\documentclass[aps,pra,twocolumn,groupedaddress,floatfix]{revtex4-1}

\usepackage{amsmath}
\usepackage{bm}
\usepackage{graphics}
\usepackage{times}
\usepackage{graphicx}
\usepackage{xcolor}
\usepackage{array,multirow,graphicx}
\usepackage{multirow,hhline,graphicx,array}
\usepackage{rotating}
\usepackage{multirow}
\usepackage{booktabs}
\usepackage[markup=underlined]{changes}
\pacs{32.80.Rm, 32.80.Wr, 32.80.Fb}

\begin{document}

\title{R-matrix with time-dependence calculations for three-sideband RABBITT in helium}
 
\author{A.~T.~Bondy$^{1,2}$}
\email{bondy11u@uwindsor.ca}
\author{J.~C.~del Valle$^{2}$}
\author{S.~Saha$^{2,3}$}
\author{K.~R.~Hamilton$^{4}$}
\author{D.~Bharti$^{5}$} 
\author{A.~Harth$^{5,6}$}
\author{K.~Bartschat$^{2}$}

\affiliation{$^1$Department of Physics, University of Windsor, Windsor, ON N9B 3P4, Canada}
\affiliation{$^2$Department of Physics and Astronomy, Drake University, Des Moines, IA 50311, USA}
\affiliation{$^3$Department of Physics, Panskura Banamali College (Autonomous), Panskura, West Bengal 721152, India}
\affiliation{$^4$Department of Physics, University of Colorado at Denver, Denver, CO 80204, USA}
\affiliation{$^5$Max-Planck-Institute for Nuclear Physics, D-69117 Heidelberg, Germany}
\affiliation{$^6$Center for Optical Technologies, Aalen University, D-73430 Aalen, Germany}

\date{\today}

\begin{abstract}
Following up on a recent paper [Bharti {\it et al.}, Phys. Rev. A~{\bf 109} (2024) 023110],
we compare the predictions from several $R$-matrix with time-dependence calculations for a modified three-sideband 
version of the ``reconstruction of attosecond beating by interference of two-photon transitions" \hbox{(RABBITT)} 
configuration applied to helium.
Except for the special case of the threshold sideband, which appears to be very sensitive to the
details of coupling to the bound Rydberg states, increasing the number of coupled states in the 
close-coupling expansion used to describe the ejected-electron--residual-ion interaction hardly changes the results.
Consequently, the remaining discrepancies between the experimental data and the 
theoretical predictions are likely due to uncertainties in the experimental parameters, particularly the detailed knowledge of the laser pulse.
\end{abstract}

\maketitle

\section{Introduction}\label{sec:Intro}

The reconstruction of attosecond beating by interference of two-photon transitions \hbox{(RABBITT)} is a widely-employed technique to characterize an attosecond pulse train and measure attosecond time delays in photo\-ionization processes, e.g.,~\cite{Paul2001,Muller2002,Klunder2011}. 
In the typical RABBITT scheme, photoionization occurs by
superposition of an infrared (IR) pulse and an extreme ultraviolet (XUV)
pulse, which is composed of odd high-order harmonics of the IR field. The
photoelectron spectra generated in this scheme contain one
sideband between every two consecutive main peaks generated by the XUV
field alone.  

In 2019, Harth~{\it et al.}~\cite{Harth2019} introduced the \hbox{3-SB RABBITT} scheme, 
a variant in which the XUV comb comprises multiple odd harmonics of the frequency-doubled
IR field. As a result, the photoelectron spectrum in \hbox{3-SB} RABBITT 
exhibits three sidebands between the main peaks. We label these peaks
$S_l$~(low), $S_c$~(center), and $S_h$~(high), respectively.

The original idea behind the scheme was to compare the RABBITT phase, extracted from 
oscillations in the ejected-electron signal as a function of the delay between the IR and the XUV comb, for the center sideband in both the three-sideband (3-SB) and single sideband (1-SB) setups. Through this comparison, one could directly obtain information about the continuum-continuum phase~$\Phi_{cc}$ that is introduced by the additional IR transitions that form the sidebands. In order to extract the
Wigner phase, which is associated with the single-photon ionization process and the corresponding photo\-ionization delay, measurements of $\Phi_{cc}$ are generally replaced by theoretical assumptions~\cite{Dahlstr_m_2012,Dahlstr_m_2014,RevModPhys.87.765}.  However, a direct measurement of the relative phase between two continuum-continuum transitions was reported by Fuchs~\emph{et al.}~\cite{Fuchs:s}. 

While a direct measurement of $\Phi_{cc}$ has not yet been achieved with the \hbox{3-SB} \hbox{vs.~1-SB} setup, in part due to
unexpected complications discovered since the original proposal, the multi-sideband setup was analyzed in more detail theoretically by Bharti~{\it et al.}~\cite{Bharti2021} with supporting
calculations for atomic hydrogen. Then the validity of the scheme was illustrated experimentally first on argon~\cite{Bharti2023} and, most recently, on helium~\cite{Bharti2024}.  Both experiments were again supported by numerical calculations, using both single-active-electron (SAE) models and the non-perturbative all-electron $R$-matrix (close-coupling) with time-dependence (RMT) approach~\cite{BROWN2020107062}. Not surprisingly, only qualitative agreement between experiment and theory was achieved for the argon target, which is still rather complex for a detailed numerical treatment.  The agreement improved, to some extent, for the helium target, but the remaining differences between the measurements and the theoretical predictions certainly warrant further investigations.

The purpose of the present follow-up paper, therefore, is a thorough analysis of the sensitivity of the theoretical predictions to changes in the model. While it is not the atomic hydrogen wished for by many theorists, the helium target is very suitable for this, since the ejected-electron--residual-ion inter\-action is effectively electron scattering from the He$^+$ target. In addition, the initial state as well as intermediate excited bound states, and even doubly-excited states, can be obtained as solutions of the ``collision problem'' with slightly modified boundary conditions, hence ensuring consistency between the one-electron target problem (He$^+$) and the two-electron bound-state and collision models. Since relativistic effects are negligible for the present purposes, and the non\-relativistic one-electron orbitals of He$^+$ are known analytically, one can perform a convergence study by increasing the number of discrete physical bound states in the close-coupling expansion and supplementing them with short-range pseudo\-states to account for coupling to high-lying Rydberg states as well as the ionization continuum.  This is the idea behind the ``convergent close-coupling'' (CCC)~\cite{PhysRevA.46.6995} and $R$-matrix with pseudo\-states (RMPS)~\cite{0953-4075-29-1-015} methods
that have been highly successful in describing electron collision and steady-state weak-field photo\-ionization processes for the past nearly three decades. 

This paper is organized as follows. We begin in Sec.~\ref{sec:Theory} with a brief review of the RMT approach, 
with specific emphasis on the models used for this paper. We then present our results in
Sec.~\ref{sec:Results} and summarize our main conclusions in Sec.~\ref{sec:Summary}.  Unless indicated
otherwise, atomic units are used throughout. 

\section{Theory}\label{sec:Theory}
We employed the general \hbox{$R$-matrix} with time dependence (RMT)
method~\cite{BROWN2020107062} to propagate the initial $(1s^2)^1S$ bound state of helium in time
under the influence of the time-dependent external electric laser field. 
In order to do so, the program first needs input from 
a time-independent code, specifically the basis into which we expand the
total wave\-function inside the $R$-matrix box and the dipole matrix elements between all
basis functions.  Furthermore, the target orbitals for the ${\rm e- He^+}$ collision problem
need to be provided.  We used the RMATRX-II code~\cite{RMAT2}, whose output has been interfaced with RMT. 

In our recent work~\cite{Bharti2024}, we set up the simplest conceivable RMT model, namely a 
non\-relativistic \hbox{1-state} approach, which we will label ``1st'' below.  As a follow-up 
on the previous study to carry out the sensitivity check of our predictions, we now define 
a \hbox{3-state} (``3st''), a \hbox{6-state} (``6st''), and finally a  
\hbox{10-state} (``10st'') model.  In addition to
just the He$^+(1s)$ ionic ground state in the 1st approach, the latter models include the excited $n\!=\!2$ (3st) and $n\!=\!3$ (6st)
physical excited states, as well as $n\!=\!4$ pseudo\-states (10st).  We used the analytically known \hbox{$1s, \ldots, 3d$}
orbitals to represent coupling between the low-lying ionic states.  On the other hand, the $n\!=\!4$ orbitals 
were taken to fall off as $\exp(-1.27\,r)$, where $r$ is the distance from the origin.  Hence,  
they have a significantly shorter range than the physical $n\!=\!4$ orbitals that fall off as $\exp(-0.5\,r)$. 
While introducing even more (pseudo)states would have been desirable, the computational complexity was 
prohibitive and, as we will see below, largely unnecessary.

While using the $1s$ orbital of He$^+$ is not optimal to obtain the best ground-state energy in the 1st model, this disadvantage is mitigated already in part by the bound-continuum and continuum-continuum terms in the \hbox{$R$-matrix} Hamiltonian. In the present extension, the situation is further improved
through the additional target orbitals associated with the excited ionic states.  In fact, the largest
improvement is provided by the $n\!=\!4$ pseudo-orbitals, which lower the theoretical ground-state energy to $-2.894$ in the 10st model, compared to $-2.873$ (1st), $-2.885$ (3st), and  $-2.887$ (6st),
respectively. While the RMT code has the option of adjusting the initial state energy to its
experimental value of $-2.9035$~\cite{NIST}, we did not use this option, since i)~we checked that the effect is negligible while ii)~the visibility of the results shown below is better without the adjustment due to the offset caused by the slightly different ionization potentials. 

\begin{figure}[t!]
\includegraphics[width=0.80\columnwidth]{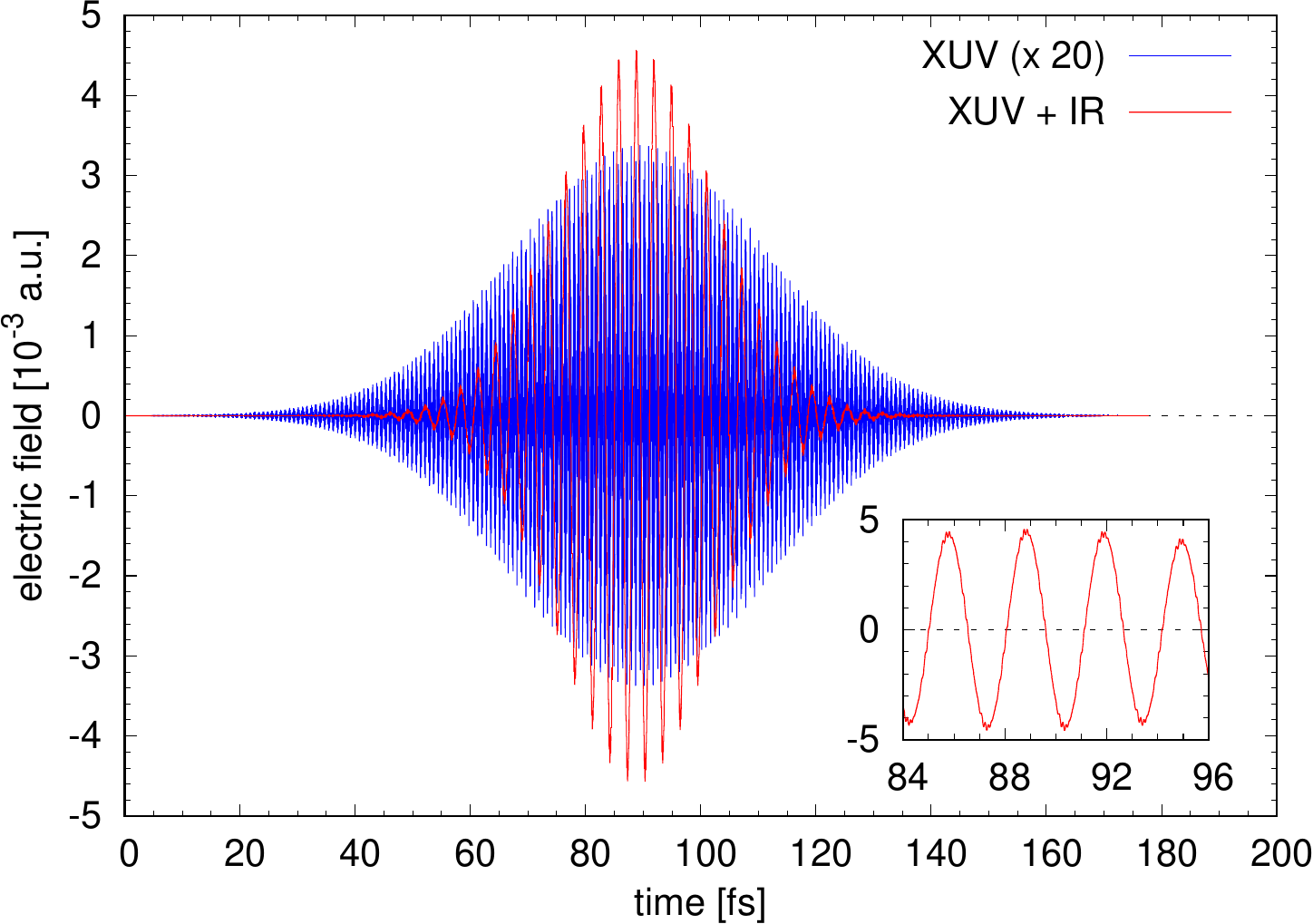}
\medskip

\includegraphics[width=0.85\columnwidth]{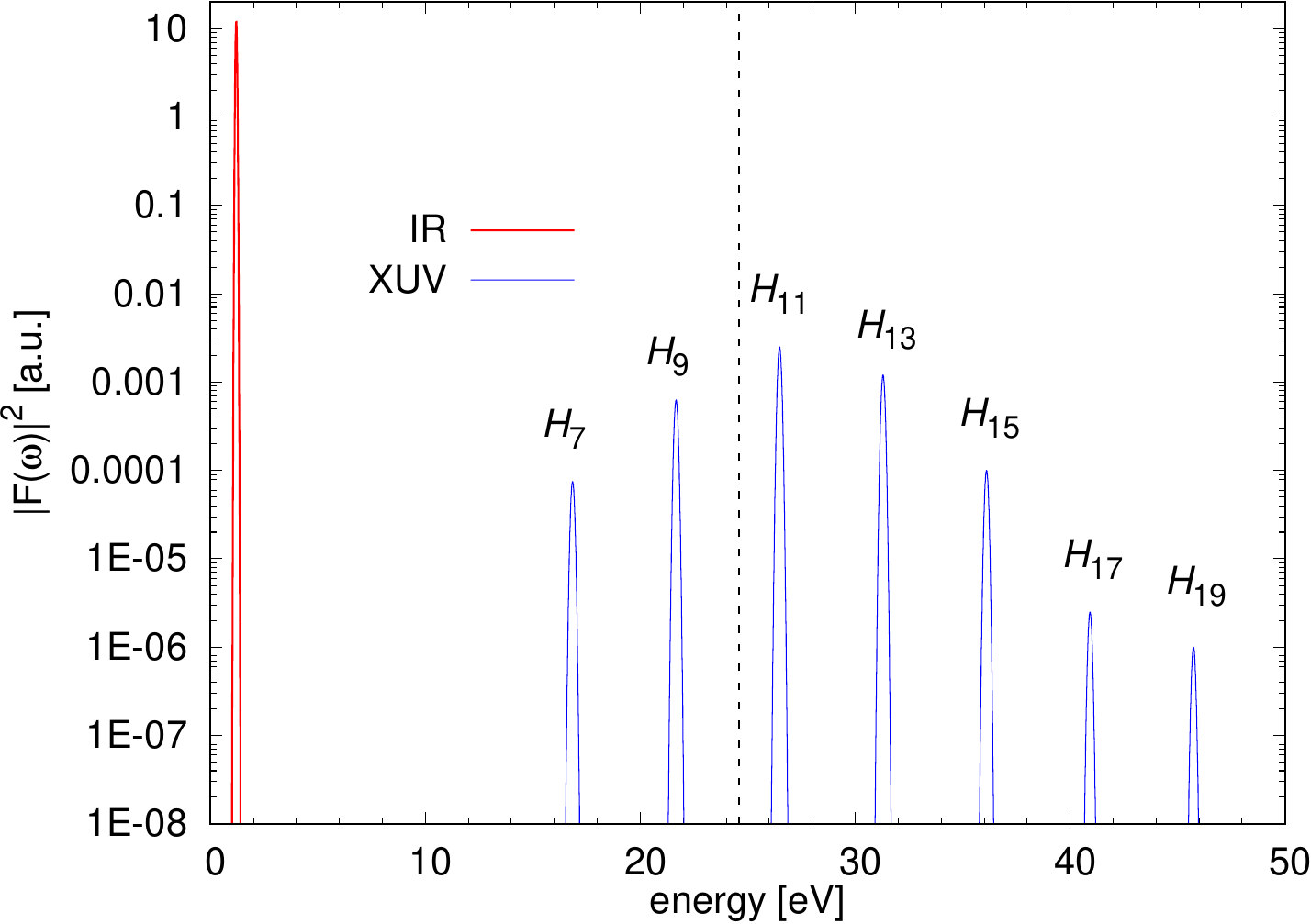}~~~~~~~
\medskip

\includegraphics[width=0.80\columnwidth]{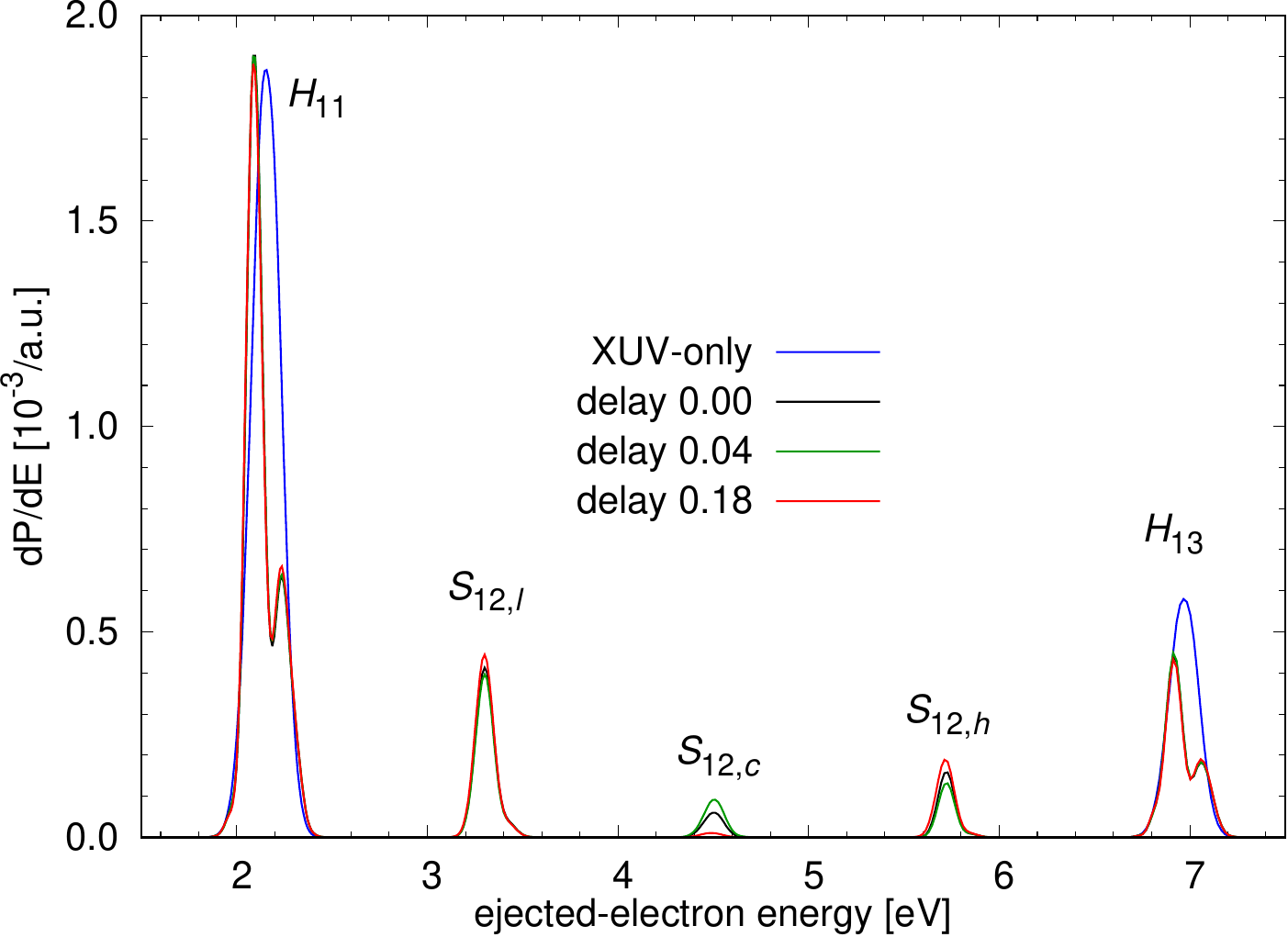}
\caption{Top: Electric field for no delay between the fundamental IR and the XUV pulse train. Note that the XUV field was multiplied by 20 for better visibility. The insert shows how small the modification of the IR field by the XUV pulse actually is. In this example, the peak intensity of the IR was set to $7 \times 10^{11}\,\rm W/cm^2$, while that of the XUV was always kept at $10^{9}\,\rm W/cm^2$.  Center: Fourier components of the fields. The XUV pulse trains contains the harmonics \hbox{$H_7,~H_9, \ldots, H_{19}$}.
The heights of the peaks were estimated based on the experimental signal~\cite{Bharti2024}.
The vertical dashed line represents the ionization threshold.
Bottom: Part of the ejected-electron spectrum. The delays are given in fractions of the fundamental IR period. See text for details. \label{fig:Fig-pulse}}
\end{figure}

Figure~\ref{fig:Fig-pulse} exhibits the pulse used in our calculations, as well as a part of the photo\-ionization spectrum.  The central wavelength of the fundamental IR was taken as 1,030~nm. 
As described by Bharti {\it et al.}~\cite{Bharti2024}, the XUV pulse train was generated as odd harmonics of the frequency-doubled fundamental radiation, i.e., starting with a wave\-length of 515~nm. 
Since an XUV spectro\-meter was not available for the experiment,
the widths and the heights of the various components shown in the center panel of Fig.~\ref{fig:Fig-pulse} were estimated based on the measured photo\-electron signal. This signal was used to generate an experimental frequency spectrum, which was then Fourier-transformed to provide the electric field as a function of time and read in on a numerical grid. 
  The uncertainty about the pulses, including a chirp in the XUV train, and this being a very challenging experiment, may be significant sources for the remaining discrepancies between experiment and theory discussed below. 

The bottom panel of Fig.~\ref{fig:Fig-pulse} shows the calculated energy-differential ionization 
probability.  For best visibility, we show it on a linear scale between the first two peaks due to the
XUV train, in this case the 11$^{\rm th}~(H_{11})$ and 13$^{\rm th}~(H_{13})$ harmonics of 515~nm. With the XUV only, there would just be two peaks in this part of the ejected-electron spectrum, while adding the fundamental IR results in the above-mentioned sidebands labelled $S_{12,l}$, $S_{12,c}$, and $S_{12,h}$, respectively.
In addition to creating sidebands, the IR is strong enough to distort the harmonic peaks. 
We clearly see that the sideband signal depends on the delay between the XUV and IR laser pulses, and also that the center sideband
promises the largest contrast.  Furthermore, due to both the different strengths of the harmonics and the energy dependence of the photo\-ionization cross section, the heights of the lower and higher sidebands are significantly different.  

Below we will show angle-integrated results from theoretical calculations for the sideband groups $S_{12}$, $S_{14}$, $S_{16}$, and $S_{18}$, as well as angle-differential results from theory and experiment for $S_{12}$, $S_{14}$. For $S_{16}$ in the angle-differential case we show theoretical predictions only since there was insufficient signal in the experiment to obtain meaningful data for the latter group. Even though the absolute heights of the peaks vary, we will see that the RABBITT phases~$\Phi_R$, which are extracted from the respective signals by fitting each of them to the functional form
\begin{equation}
S(\tau,\theta)  =  A(\theta) + B(\theta)\, \cos(4\,\omega\tau -\Phi_{R}(\theta)), 
\label{eq:3Sbsiggnal}
\end{equation}
largely follow the predictions from the decomposition approximation described in Ref.~\cite{Bharti2021}.
In Eq.~(\ref{eq:3Sbsiggnal}), $\tau$ is the delay between the IR and the XUV, while $A$ is the average signal and $B$ is the amplitude of the oscillation. These parameters depend on the energy of the ejected electron (i.e., the specific sideband of interest), and one can do the fitting with the angle-integrated signal or pick a particular angle~$\theta$ or angular range.

\section{Results and Discussion}\label{sec:Results}
\subsection{Angle-Integrated RABBITT Phase}\label{subsec:angint}
\begin{figure}[t!] 
\includegraphics[width=0.99\columnwidth]{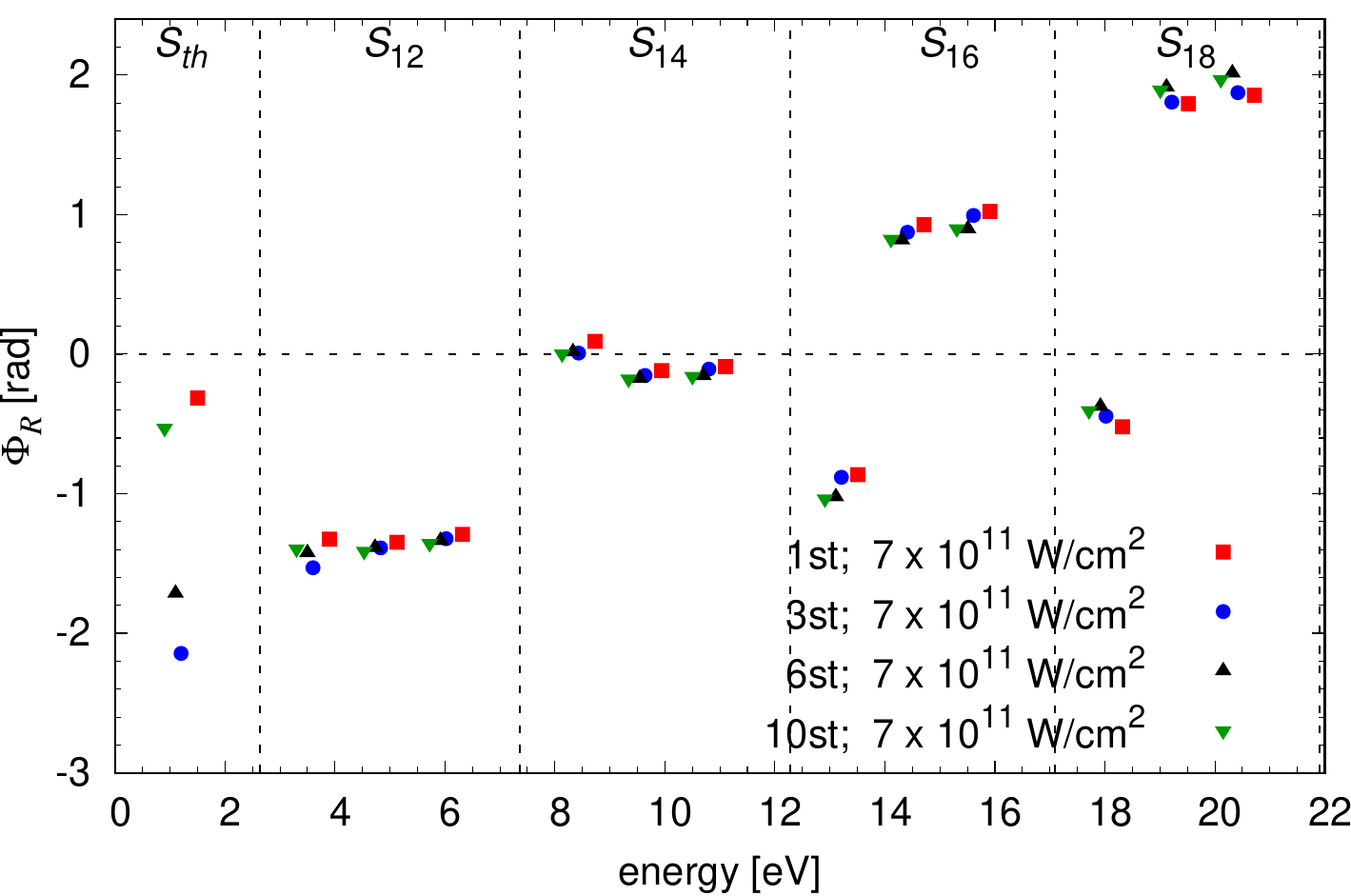}

\medskip

\includegraphics[width=0.99\columnwidth]{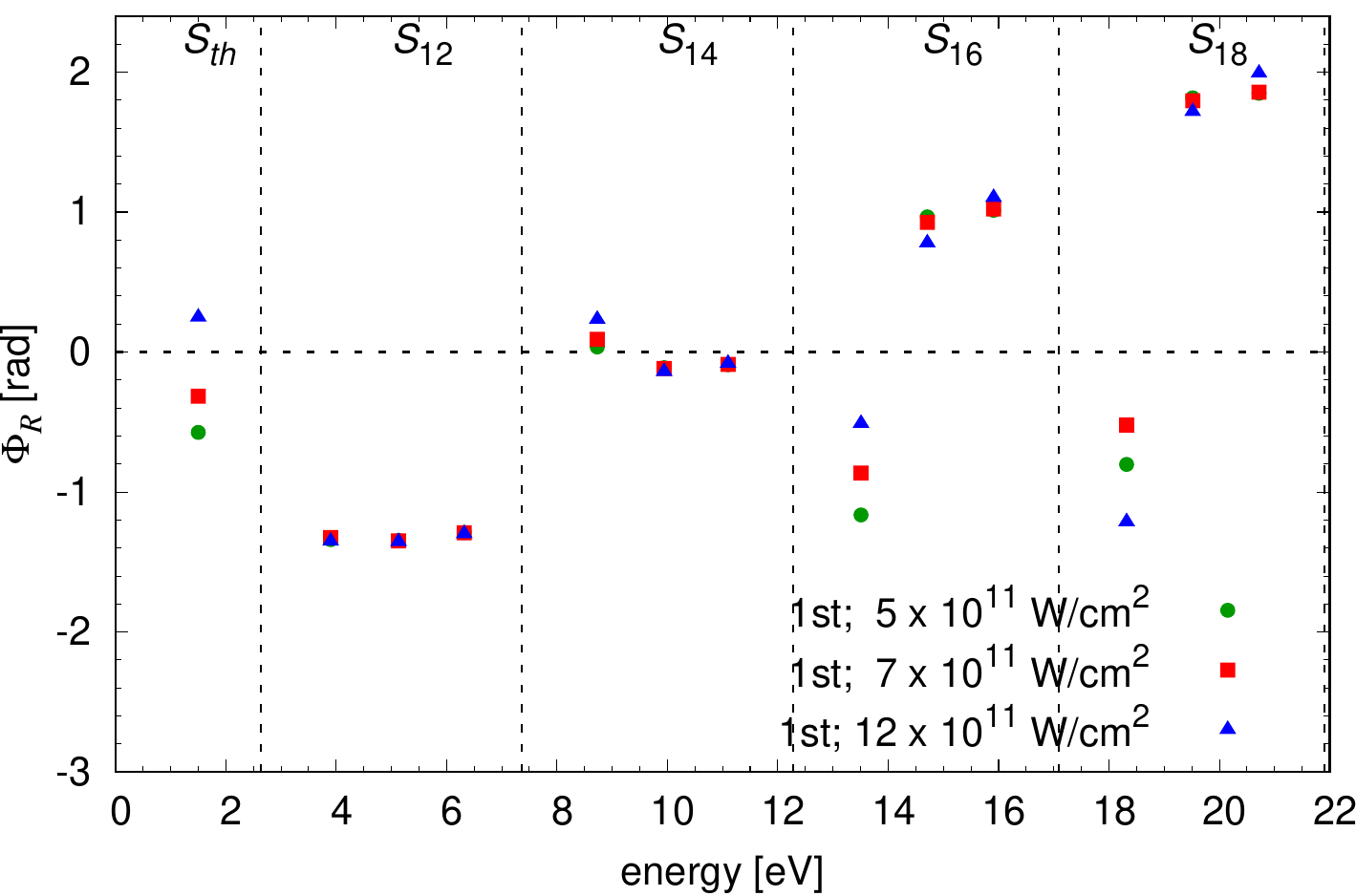}

\caption{Top: Absolute angle-integrated \hbox{RABBITT} phases, as obtained in the RMT models with a different number of coupled states for a peak intensity of
$7 \times 10^{11}\,$W/cm$^2$.  The vertical dashed lines indicate the approximate positions of the XUV harmonics to guide the eye to the sideband groups. Bottom:  Absolute angle-integrated \hbox{RABBITT} phases, as obtained in the RMT-1st model for three different intensities. 
} 
\label{fig:Phi-integrated}
\end{figure}

Figure~\ref{fig:Phi-integrated} shows our results for the RABBITT phase~$\Phi_R$ extracted from the angle-integrated ionization signals.  The top panel of the graph is for an IR peak intensity of 
$7 \times 10^{11}\,\rm W/cm^2$, as obtained in the 1st, 3st, 6st, and 10st models. The points are not
at the same energy due to the theoretical binding energy being different in each model, with the smallest 
binding energy found in the 1st model and the largest in the 10st model.  Consequently, the latter has the peaks in the 
ejected-energy spectrum at the lowest energy in each group.

There are a number of interesting observations to be made.  To begin with, except for the threshold sideband $S_{th}$, which is the higher sideband of the $S_{10}$ group and the only one above the ionization threshold,
we see only a minor dependence on the number of states included in 
the close-coupling expansion.  The main reason for the model sensitivity of the $S_{th}$ RABBITT phase appears to be
the fact that this sideband is affected by the XUV-harmonic $H_9$, which moves the active electron into a virtual or possibly even a physical state of the Rydberg series.  Hence, only a small change
in the theoretical model, which will change the ionization potential and the position of the
Rydberg states, can have a significant effect on the coupling in the so-called ``under-threshold RABBITT'' scenario.  
Experimentally, this sideband also presents exceptional challenges.  These are currently being addressed in a follow-up experiment.
We expect the results to be reported in a future publication.

\begin{figure*}[!t]
\includegraphics[width=0.66\columnwidth]{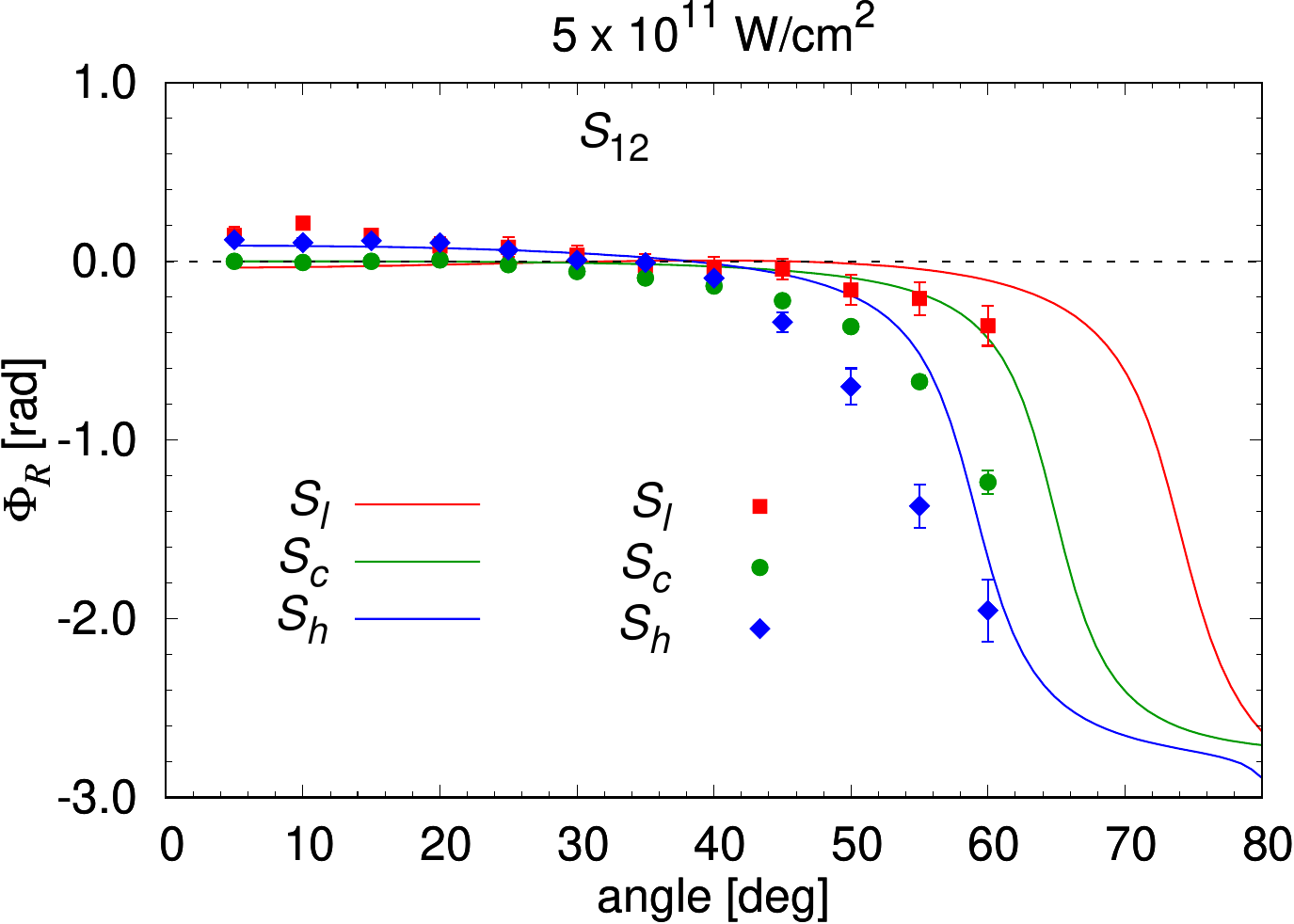}
\includegraphics[width=0.66\columnwidth]{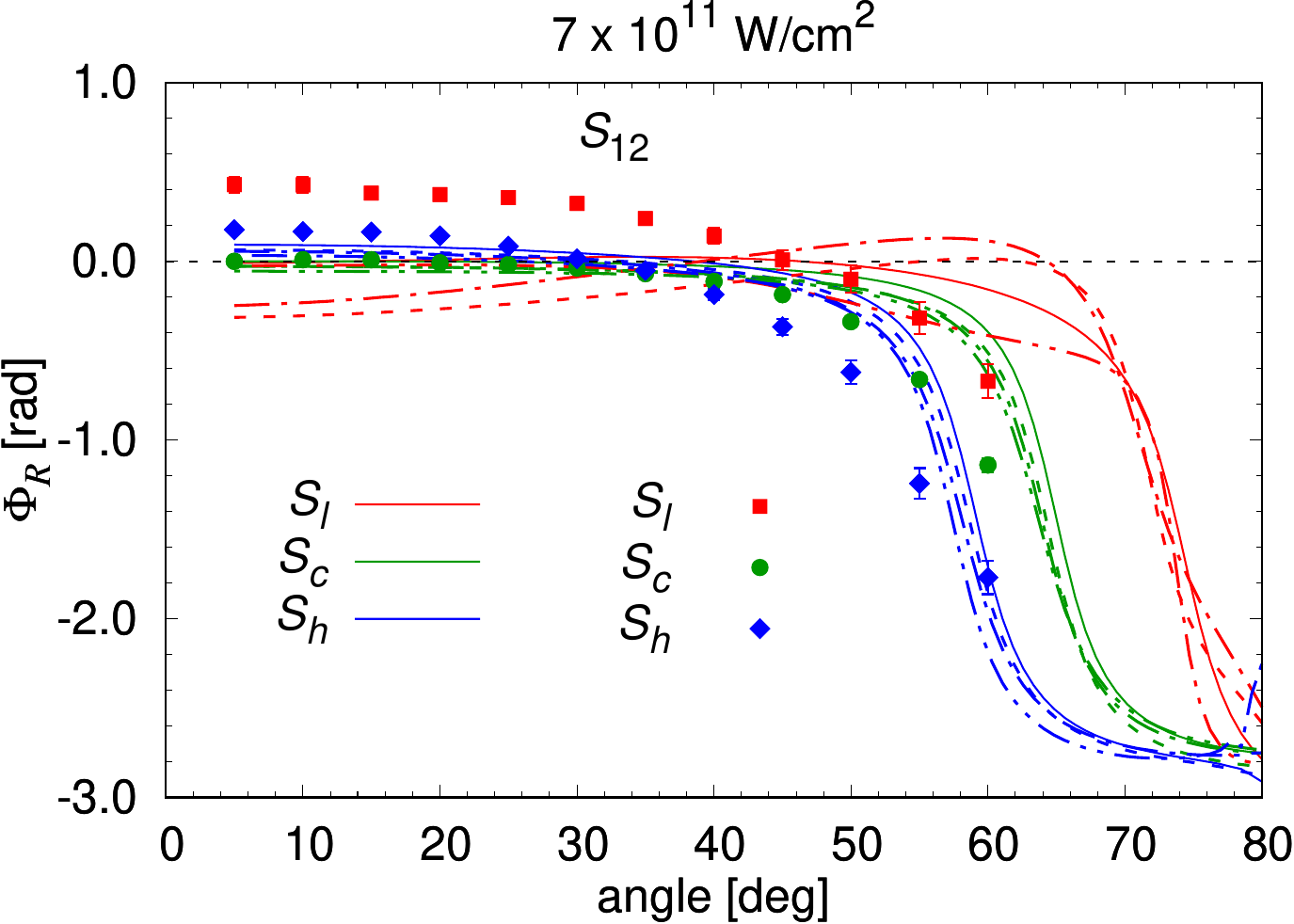}
\includegraphics[width=0.66\columnwidth]{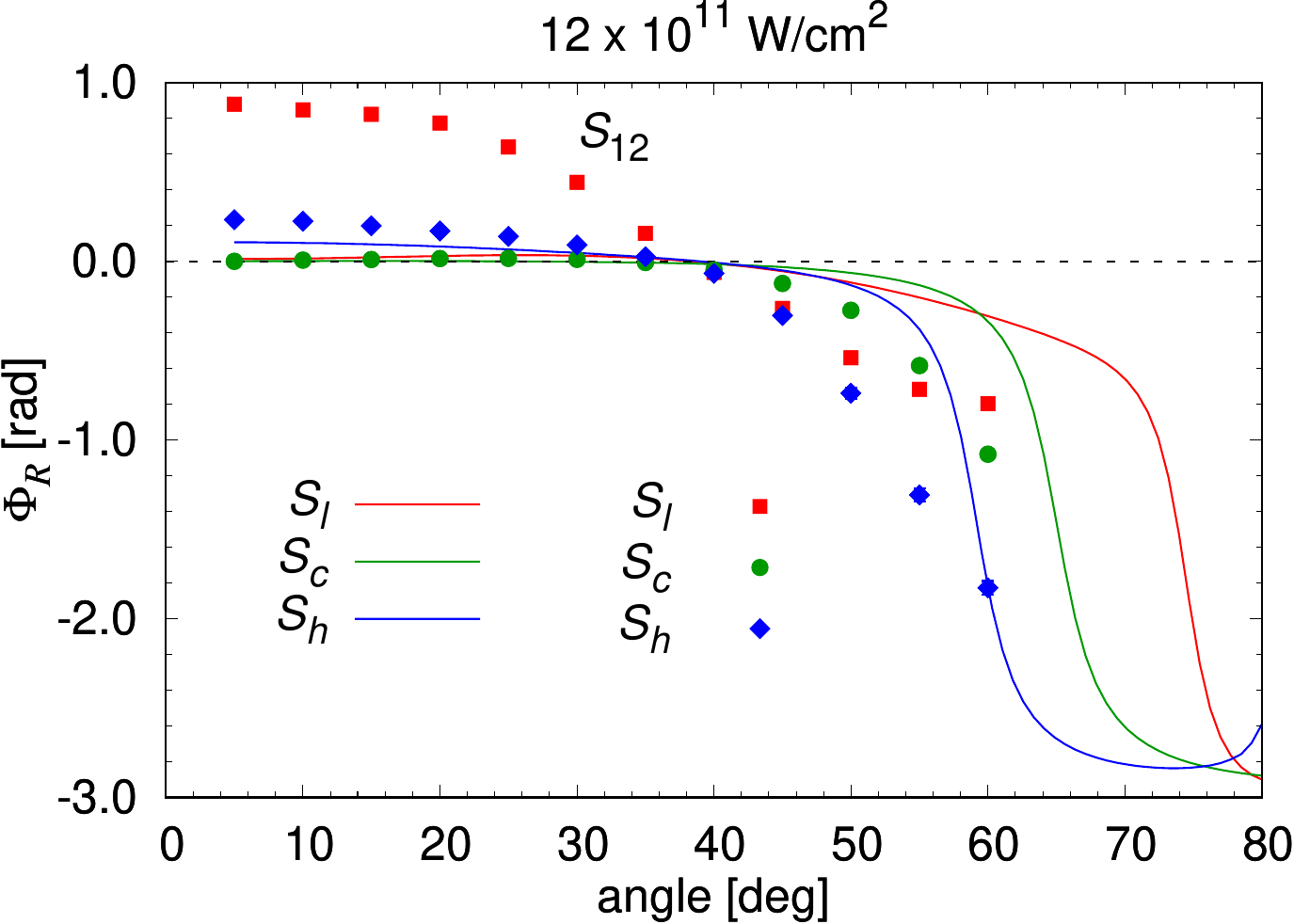}

\medskip

\includegraphics[width=0.66\columnwidth]{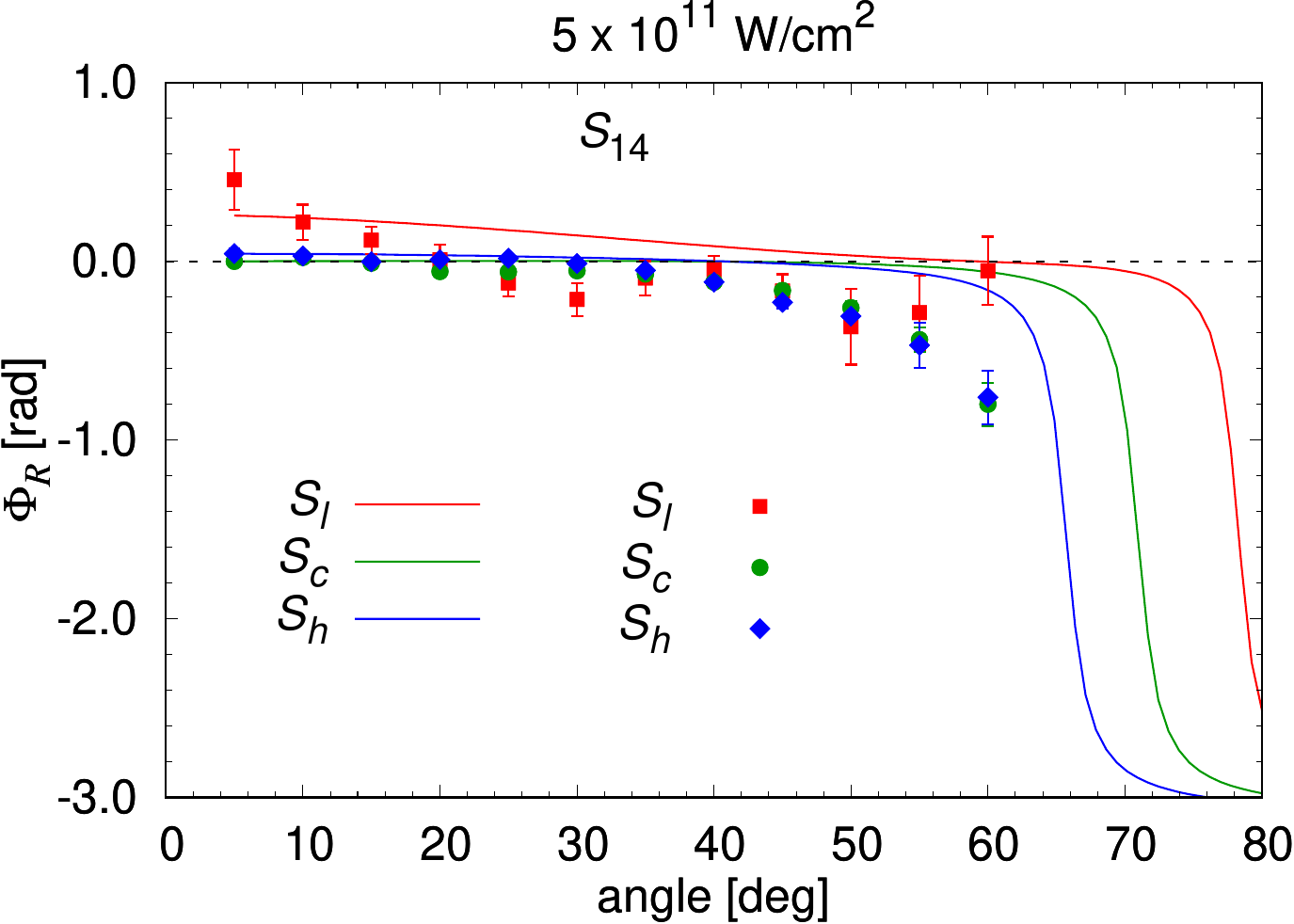}
\includegraphics[width=0.66\columnwidth]{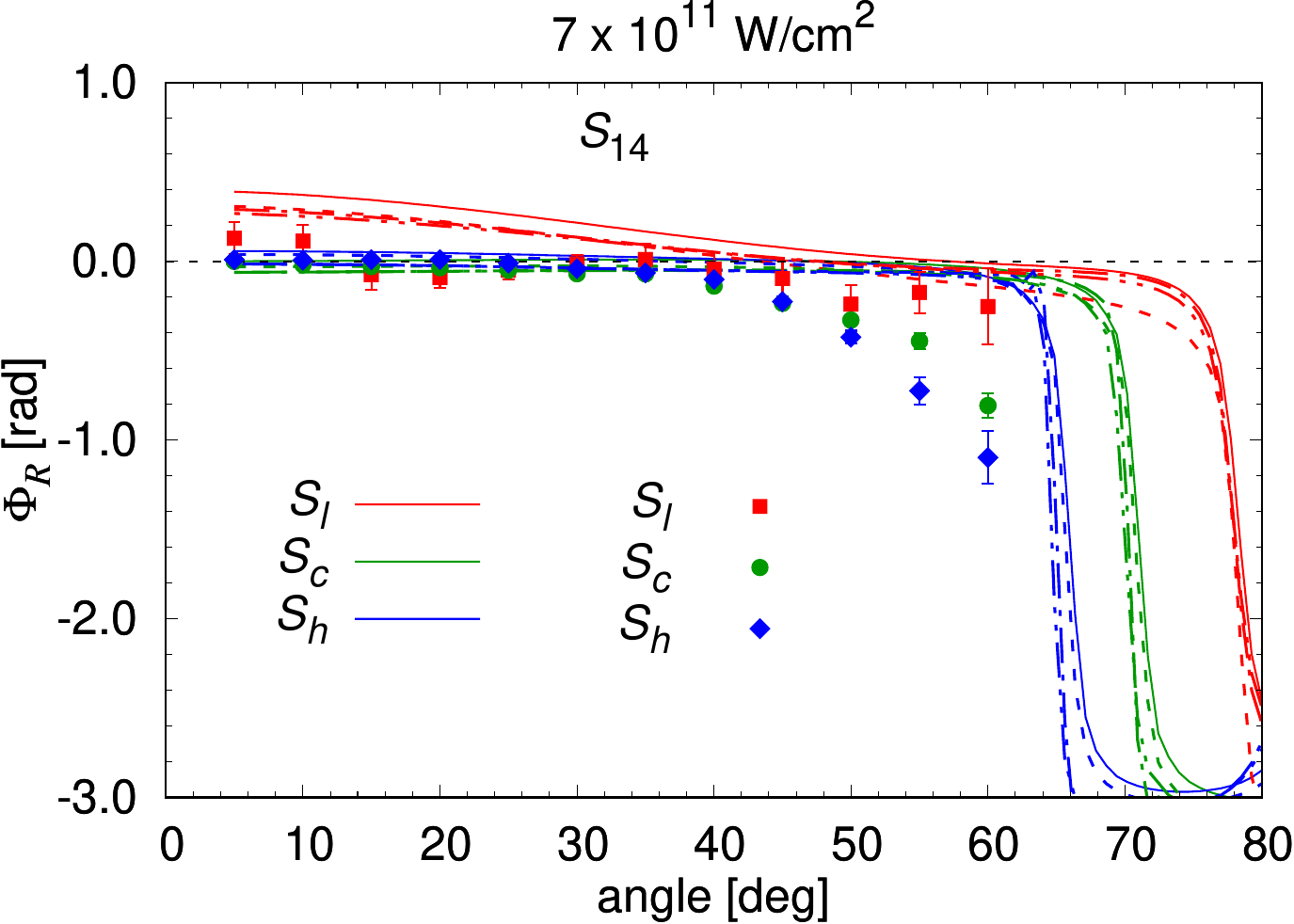}
\includegraphics[width=0.66\columnwidth]{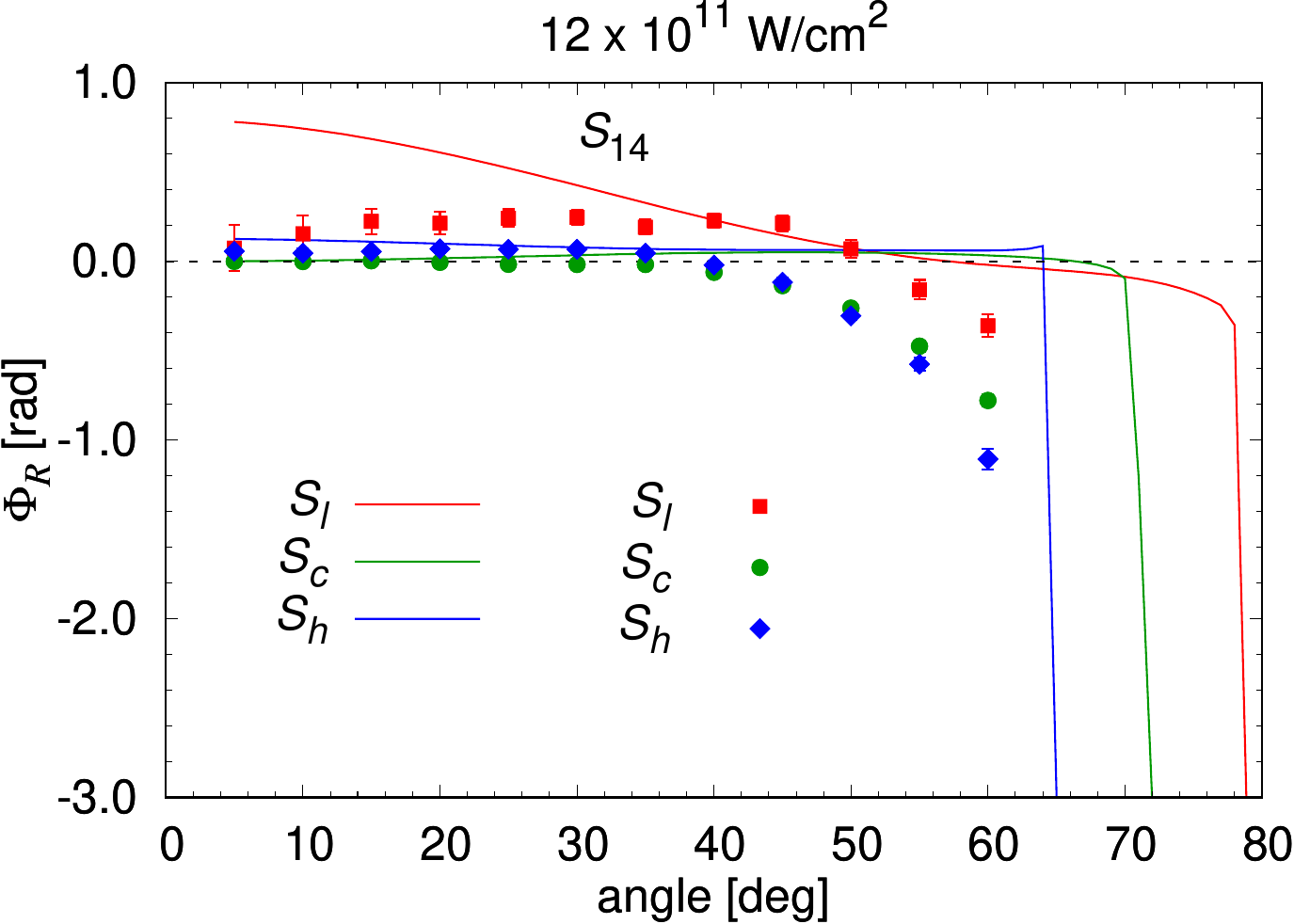}
\caption{Angle-differential results for the RABBITT phase in comparison with experiment. The experimental data are represented by the symbols, while the lines representing the theoretical predictions are distinguishable by their color and by the first one to drop rapidly with increasing angle is representing $S_h$, followed by $S_c$, and lastly~$S_l$.
For the peak intensities
of $5 \times 10^{11}\,$W/cm$^2$ and $12 \times 10^{11}\,$W/cm$^2$, we only show the 
1st results.  For $7 \times 10^{11}\,$W/cm$^2$, we add RMT predictions from
models with a different number of coupled states. The linetypes are: solid (1st),
dashed (3st), dash-dot (6st), dash-dot-dot (10st).  Due to the unknown absolute phase in the experiment,
$\Phi_R$ for the center sideband is offset to start at zero. Also, to simulate the experimental situation, a $10^\circ$ window ($\pm 5^\circ$ about the nominal angle) was integrated over.}
\label{fig:angleall}
\end{figure*}

Moving on to the sideband groups $S_{12}$ and $S_{14}$, we essentially see a confirmation of the 
``decomposition approximation''~\cite{Bharti2021}, which predicts the RABBITT phases in each of the three sidebands of a group to be identical, except for a trivial phase~$\pi$ between the center and the lower/higher sideband, which has been compensated for in the graph to show the resemblance clearly.  
Indeed, the approximation is fulfilled very well.  On the other hand, the
RABBITT phases for the lower sidebands in both the $S_{16}$ and $S_{18}$ groups differ significantly (by about two radians) from those for the center and higher sidebands.  While we do not have an explanation for the size of the difference, the most likely reason for the discrepancy from the predicted similarity of all three phases is the fact that the lower sideband is also coupled to a relatively strong next-lower rather than just the adjacent harmonic in the XUV comb.  This effect is further enhanced, albeit only slightly, by the decreasing cross section with increasing ejected-electron energy.  We emphasize that this effect was also seen in the experimental data presented by Bharti {\it et al}.~\cite{Bharti2024}.  We do not show these experimental data here, because we can extract the RABBITT phases from absolute delays between the IR and the XUV train.  These absolute delays are known theoretically from our construction of the pulses, whereas there is a free offset in the presentation of the experimental data.  The latter would reduce the visibility of the results and create a potential source of confusion if other groups were to try similar calculations as those presented here.

Given that the results for an IR peak intensity of \hbox{$7 \times 10^{11}\,\rm W/cm^2$} generally showed a small dependence on the number of states in the close-coupling expansion, and RABBITT calculations are computationally expensive due to the many delays that need to be scanned through, we only show results from the 1st model in the bottom panel of Fig.~\ref{fig:Phi-integrated}, but this time for three different IR peak intensities, namely $5 \times 10^{11}\,\rm W/cm^2$, $7 \times 10^{11}\,\rm W/cm^2$, and $12 \times 10^{11}\,\rm W/cm^2$, respectively.  The results confirm the general conclusions drawn above. In the $S_{12}$ and $S_{14}$ sideband groups, as well as the center and higher sidebands of the $S_{16}$ and $S_{18}$ groups, where the decomposition approximation seems to be fulfilled very well, there is also only a weak intensity dependence in the extracted RABBITT phase.  On the other hand, a much stronger dependence on the IR intensity is seen in the threshold sideband and also in the lower members of the $S_{16}$ and $S_{18}$ groups. The former result can be understood from the fact that the ponderomotive potential will couple the Rydberg spectrum in a slightly different way to the formation of $S_{th}$ if the IR intensity is varied, and the same coupling to the harmonics of very different strength that caused the unusual behavior in $S_{16,l}$ and $S_{18,l}$ for \hbox{$7 \times 10^{11}\,\rm W/cm^2$} will likely also be the reason at these intensities. 

\begin{figure*}
\includegraphics[width=0.490\columnwidth]{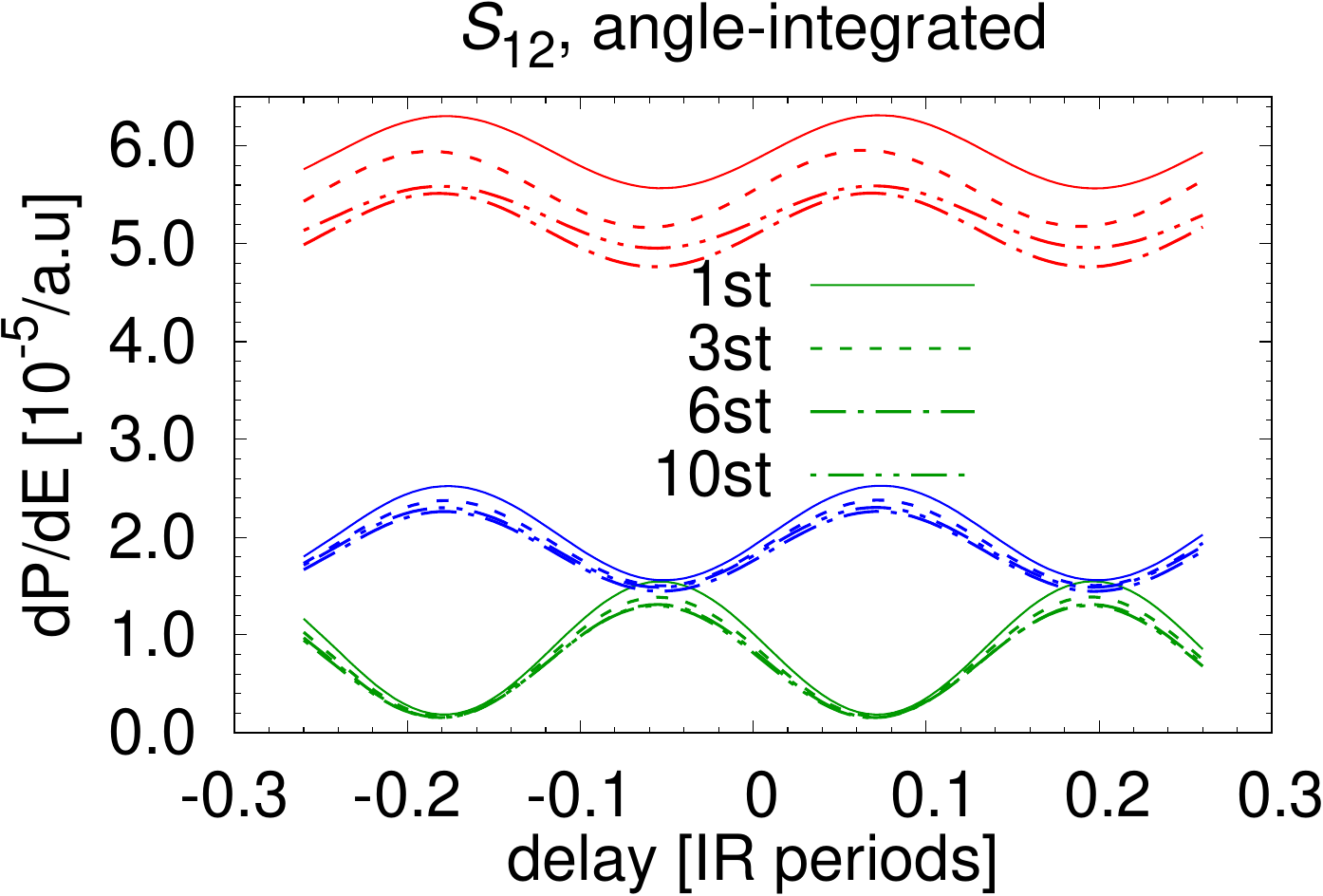}
\includegraphics[width=0.490\columnwidth]{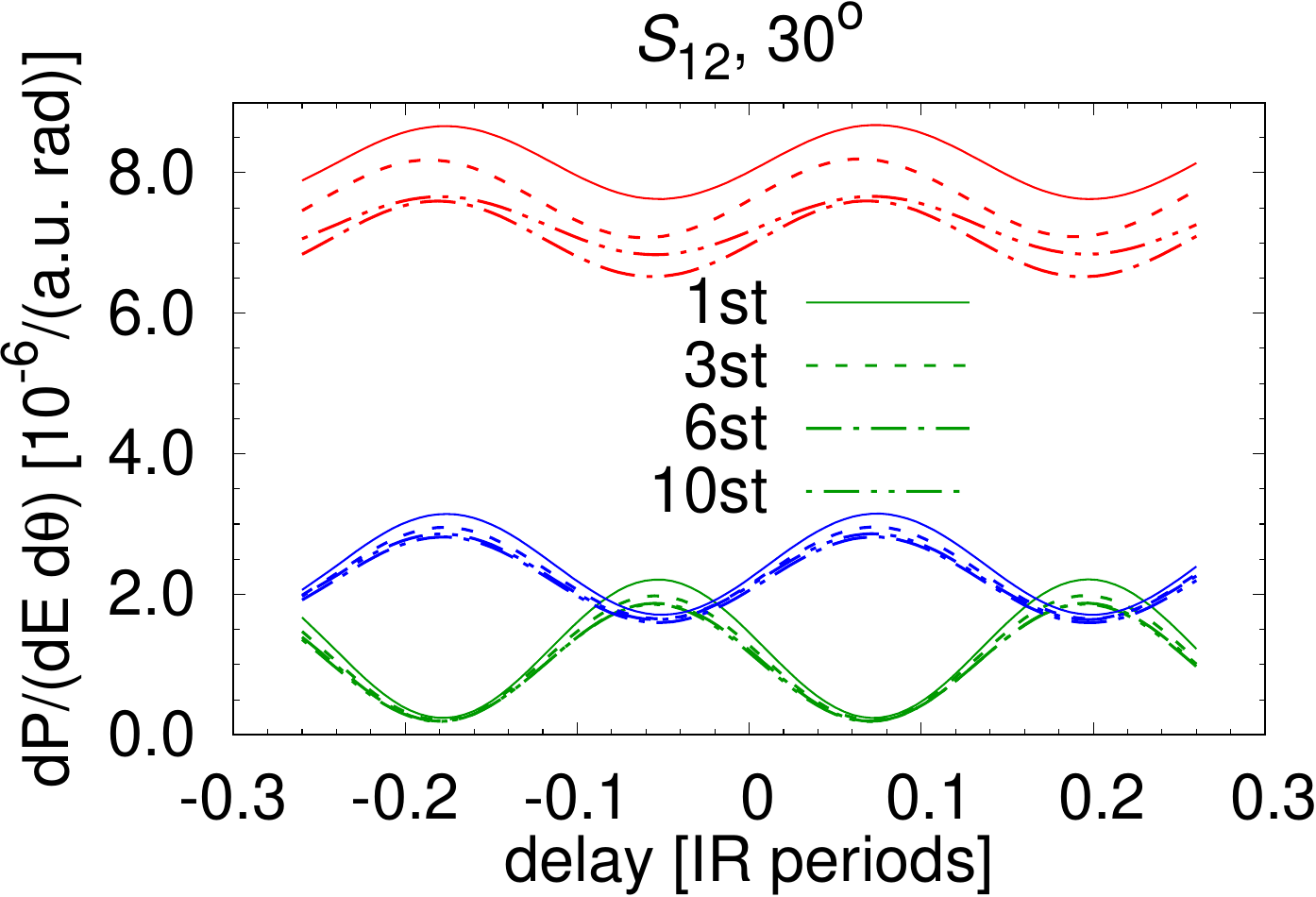}
\includegraphics[width=0.490\columnwidth]{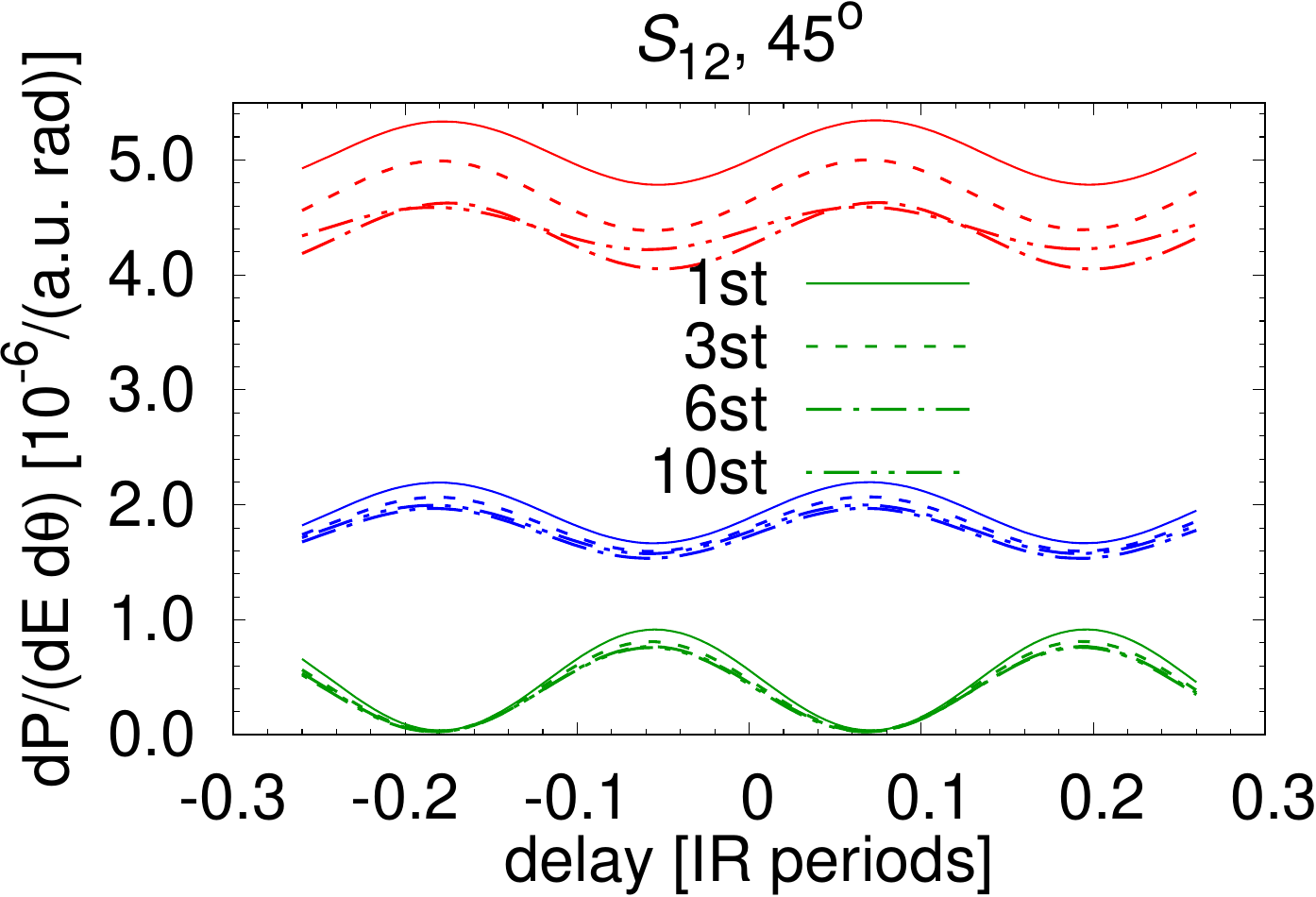}
\includegraphics[width=0.490\columnwidth]{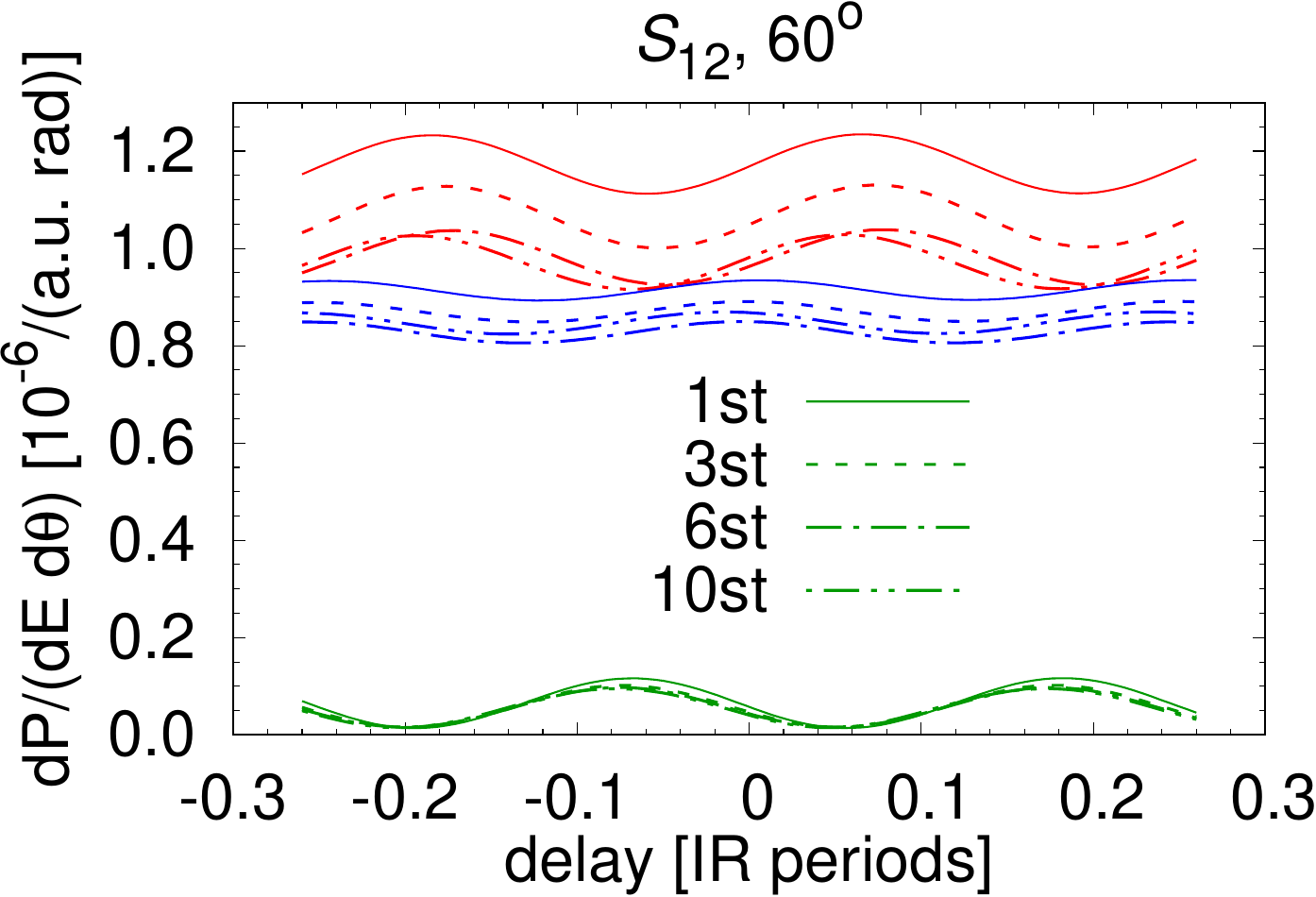}
 
\medskip

\includegraphics[width=0.490\columnwidth]{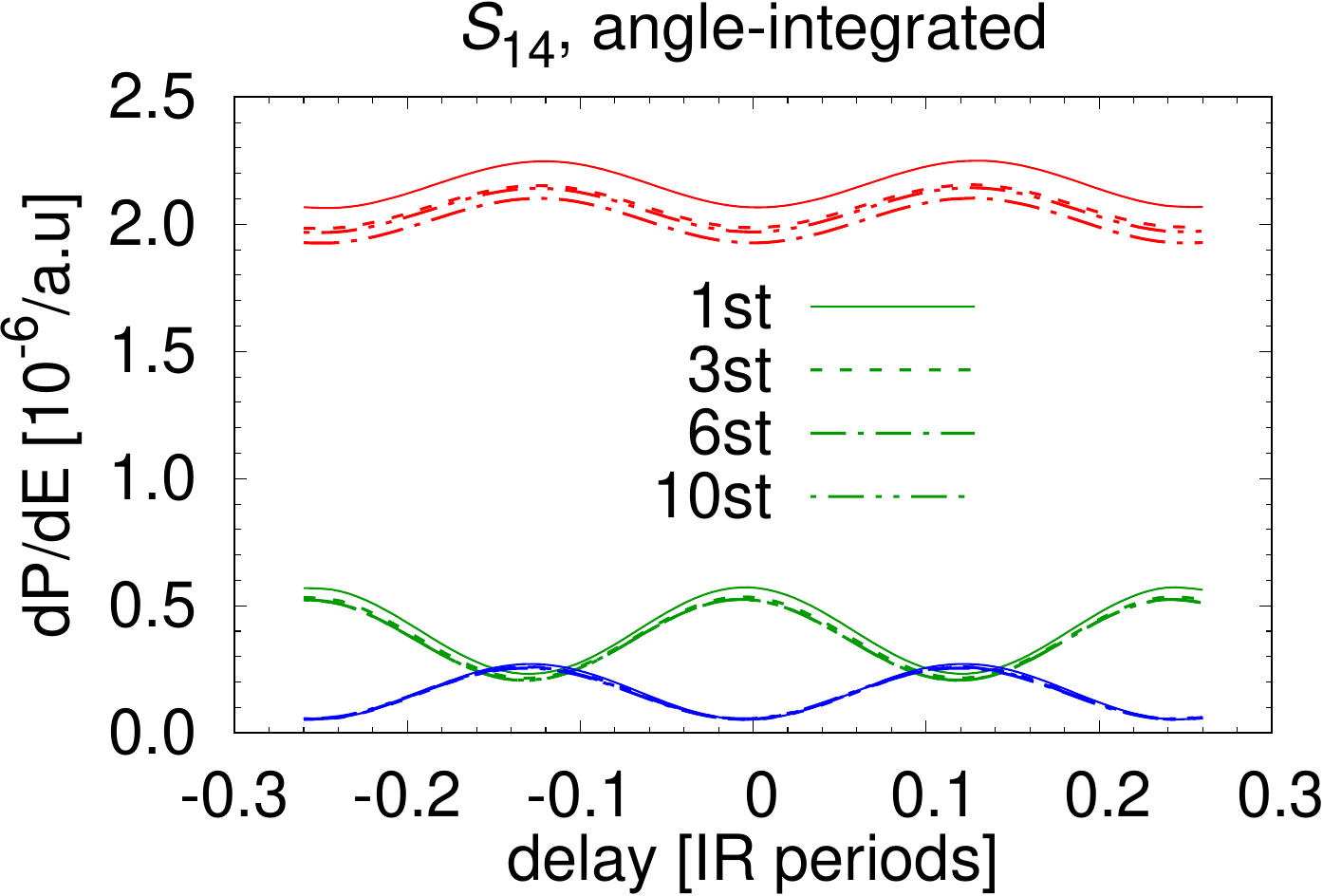}
\includegraphics[width=0.490\columnwidth]{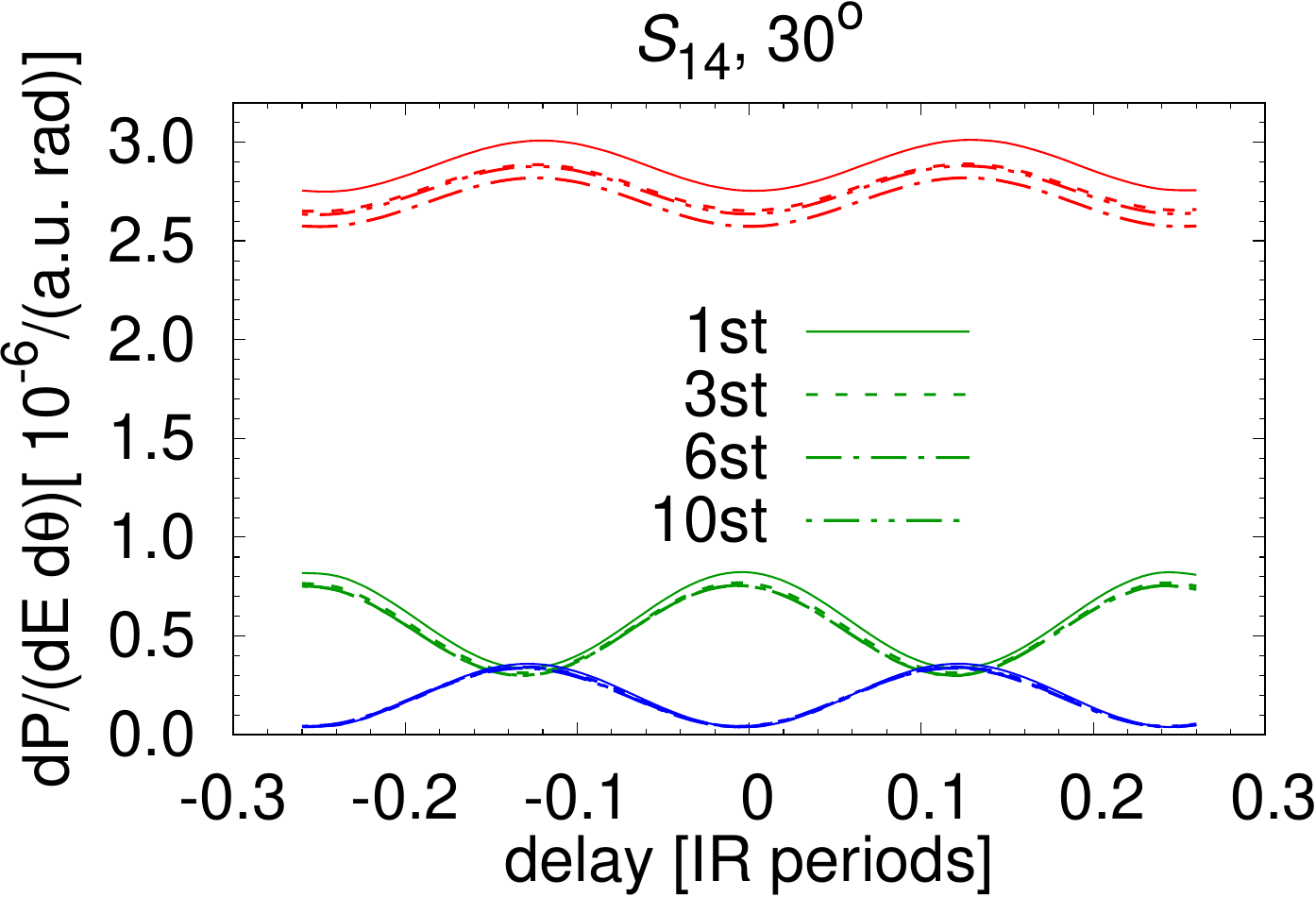}
\includegraphics[width=0.490\columnwidth]{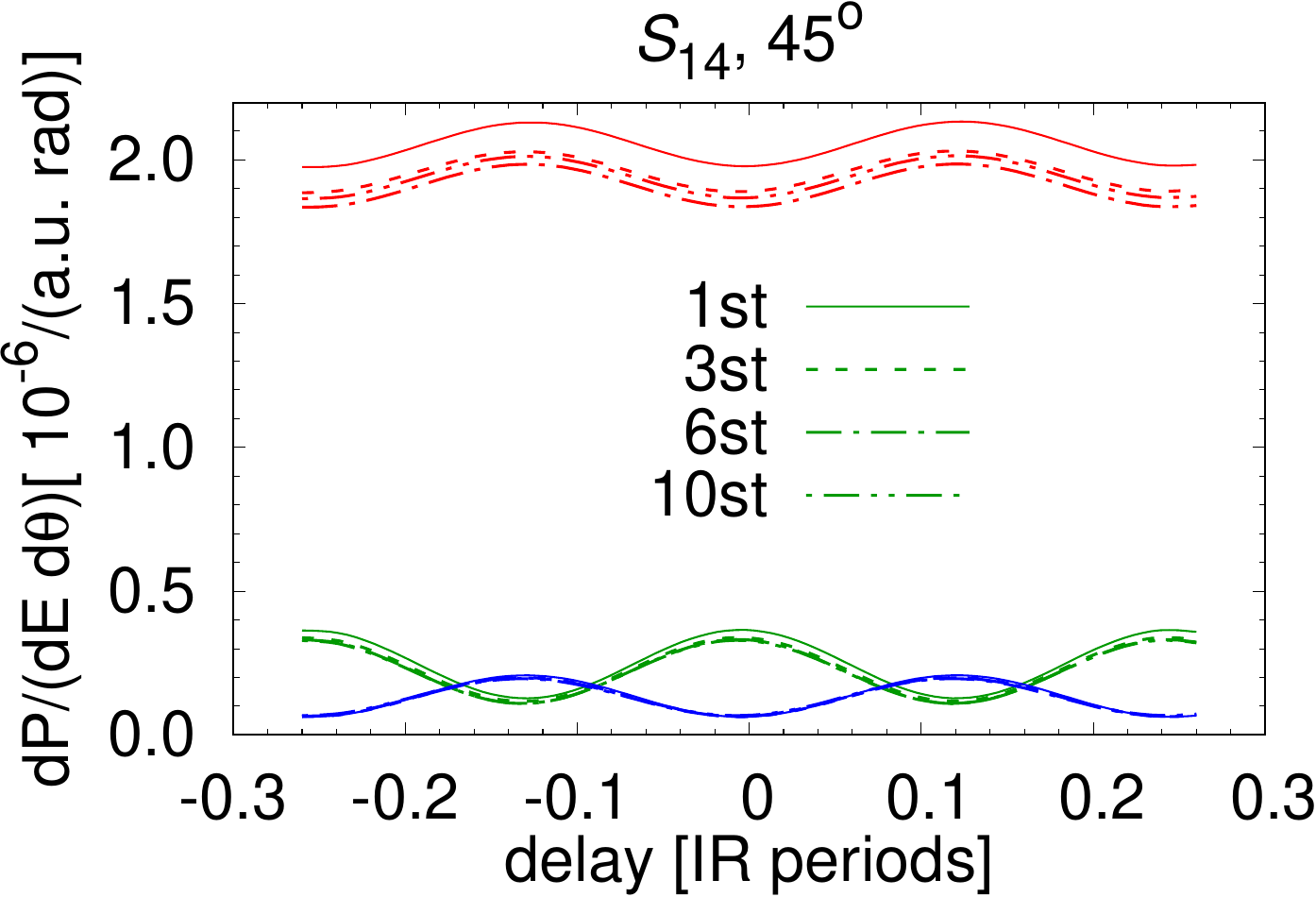}
\includegraphics[width=0.490\columnwidth]{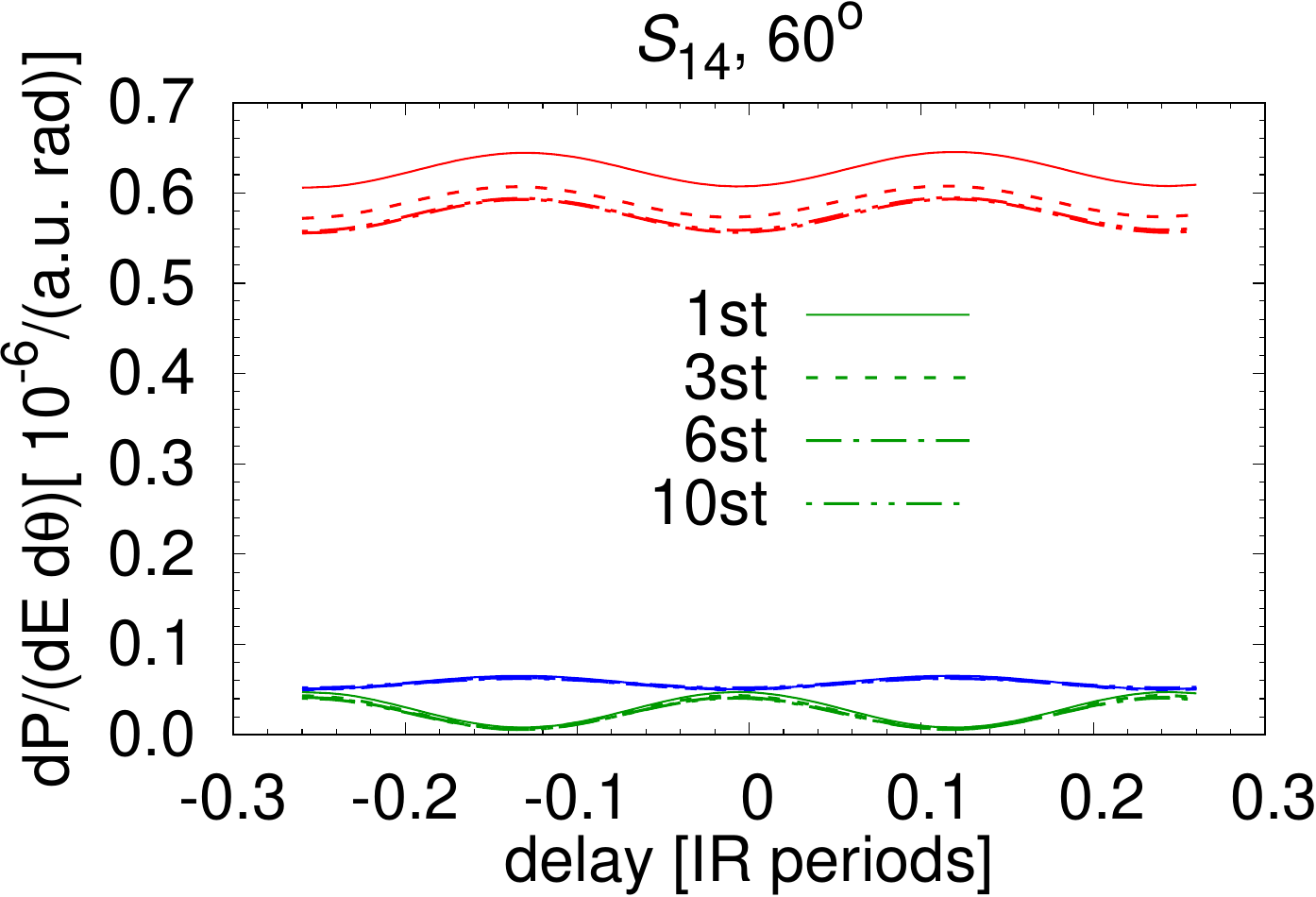}

\medskip

\includegraphics[width=0.490\columnwidth]{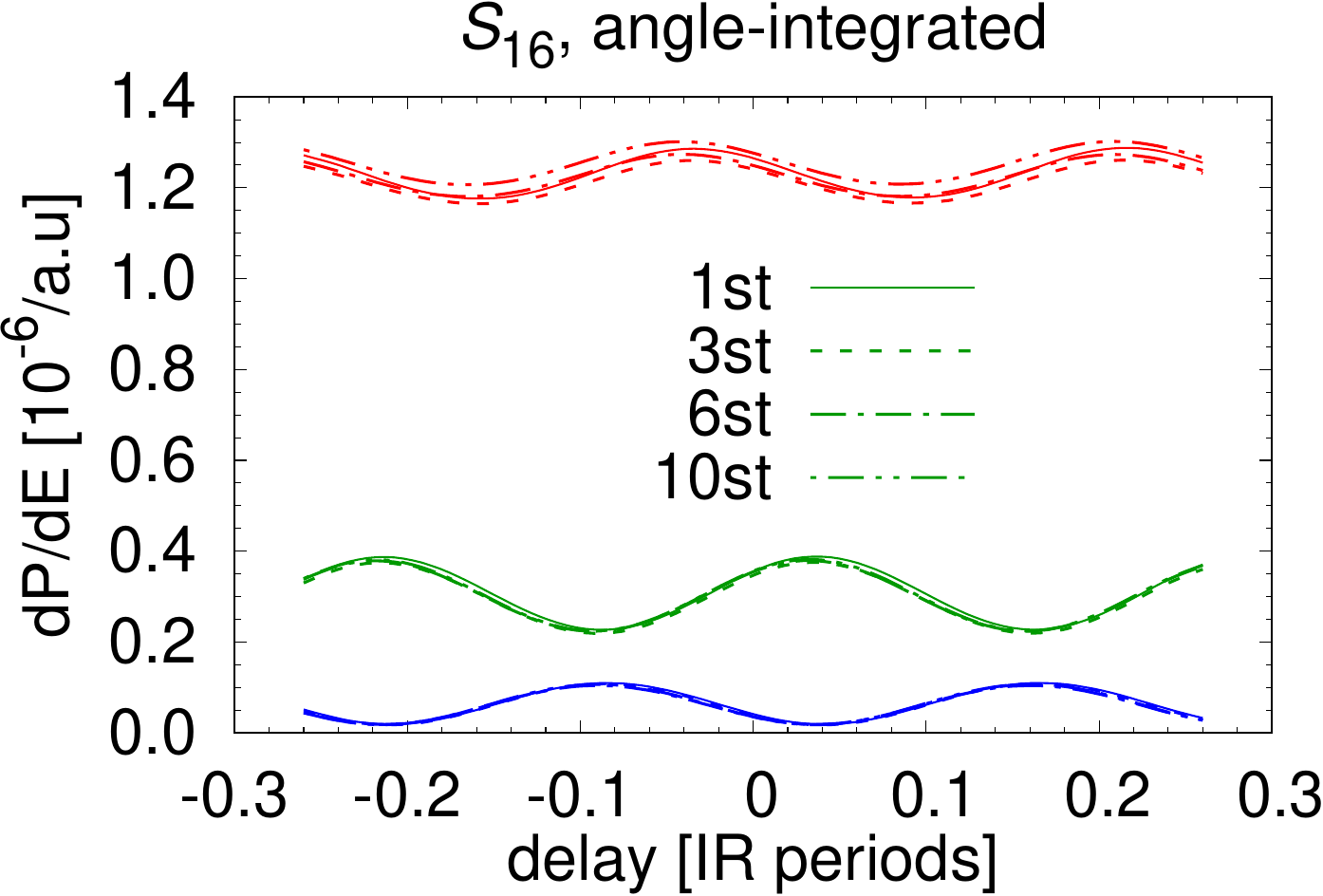}
\includegraphics[width=0.490\columnwidth]{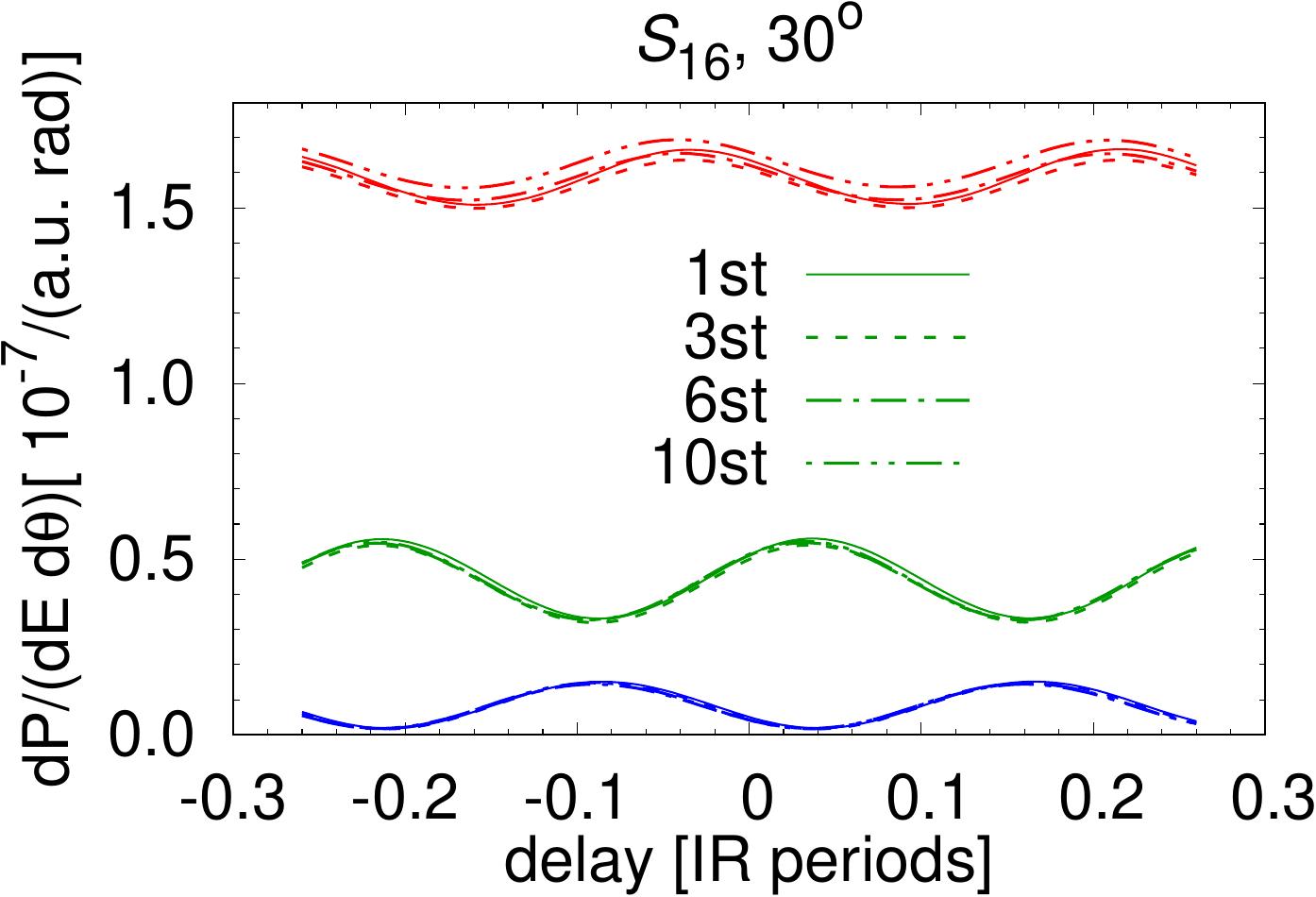}
\includegraphics[width=0.490\columnwidth]{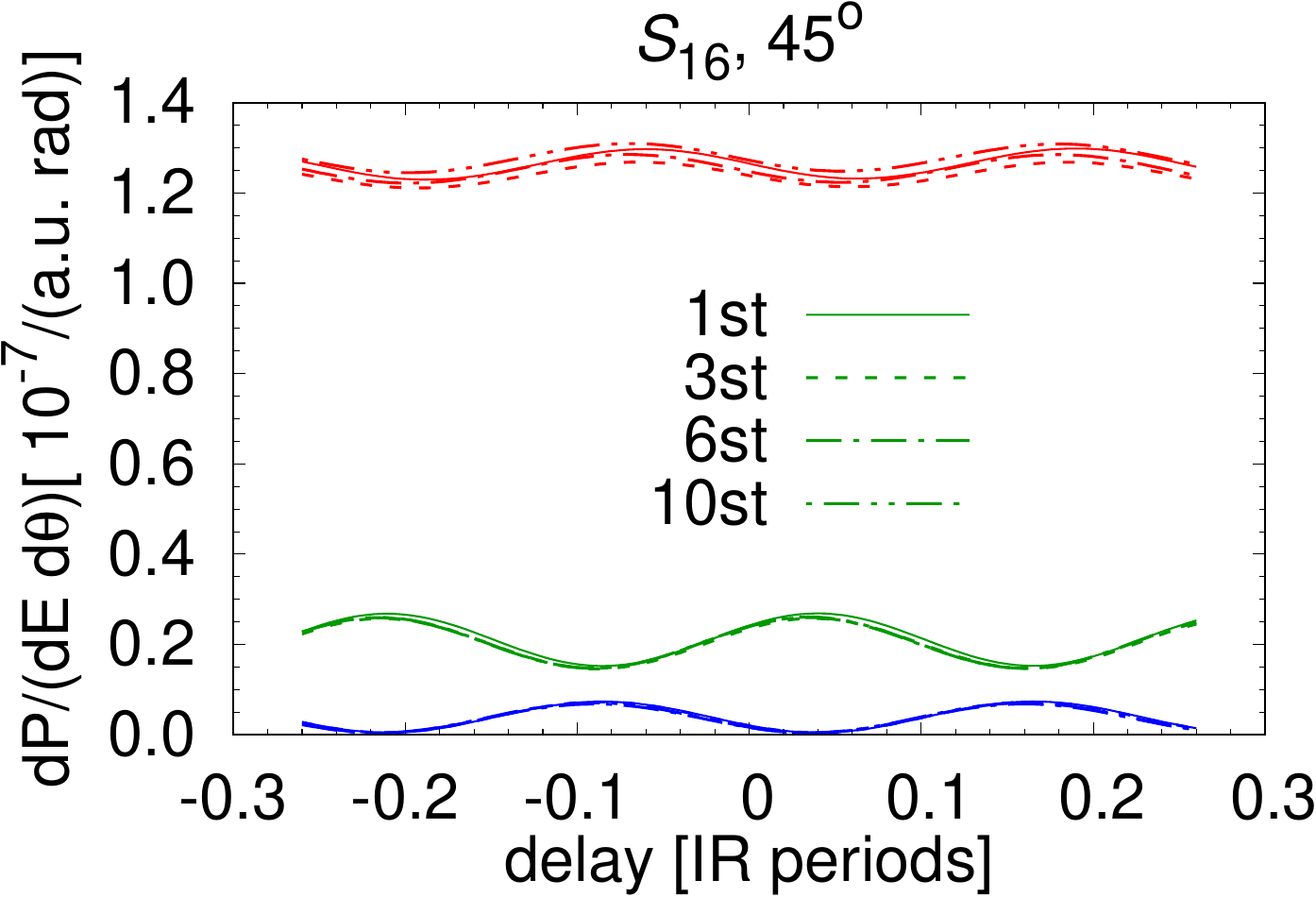}
\includegraphics[width=0.490\columnwidth]{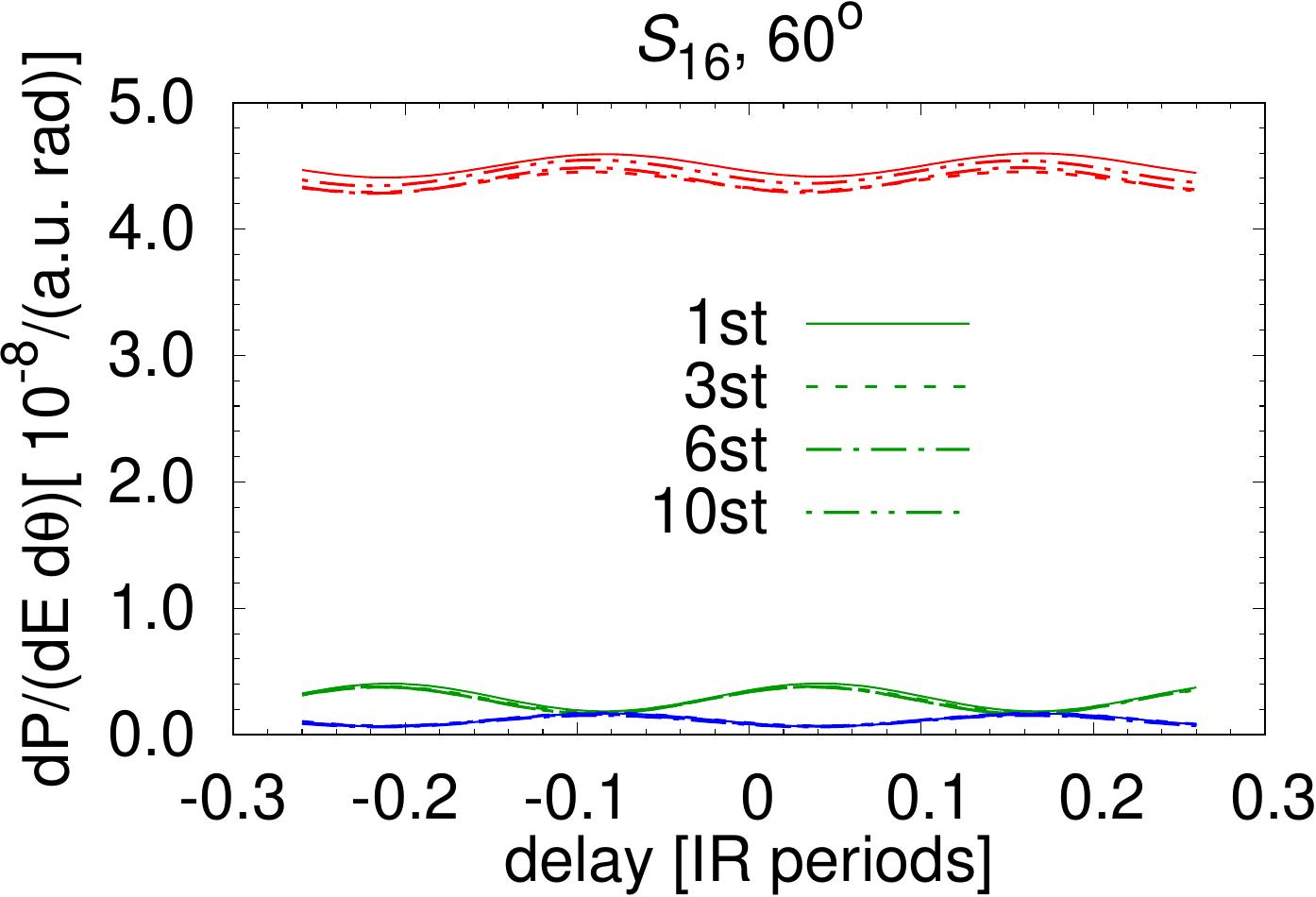}
\caption{Predicted signal oscillations for a peak intensity of 
$\rm 7 \times 10^{11}W/cm^2$ in the $S_{12}$, $S_{14}$, and
$S_{16}$ sideband groups.  The left column shows the oscillations in 
the angle-integrated case, while the other columns
are for selected ejection angles of $30^\circ$, $45^\circ$, and $60^\circ$, respectively.
Each panel contains 1-state, 3-state, 6-state, and 10-state results to illustrate
the sensitivity of the predictions to changes in the model. 
In all panels, the lower sideband (in red) has the largest magnitude.   
In $S_{12}$, the higher sideband (in blue) is the second-largest, while the
center sideband (in green) is the second-largest in $S_{14}$ and $S_{16}$, except at $60^\circ$
in $S_{14}$.
}
\label{fig:angleselected}
\end{figure*}

\subsection{Angle-Differential RABBITT Phase}\label{subsec:angdiff}
We now move on to the angle-differential case.  Figure~\ref{fig:angleall} shows
the experimental results~\cite{Bharti2024} in comparison with theoretical predictions for
IR peak intensities of $5 \times 10^{11}\,$W/cm$^2$,  $7 \times 10^{11}\,$W/cm$^2$, and $12 \times 10^{11}\,$W/cm$^2$, respectively.  We concentrate on the 
$S_{12}$ and $S_{14}$ sideband groups, since experimental data are available for these cases.

We first note that, as in the angle-integrated case, there is only a small dependence on the
number of states in the close-coupling expansion.  Consequently, the theoretical predictions
are likely converged to a higher extent than the remaining differences between the experimental data
and the corresponding prediction.  Regarding the overall trend, there is qualitative agreement 
between the data and the RMT results in that the RABBITT phase for the higher sideband in each
group drops first with increasing detection angle, followed by the phase for the center and, finally, the one for the lower sideband.  
However, the drop in $\Phi_R$ (by about $\pi$) occurs over a much narrower angular
range in the theoretical curves compared to experiment.  
It is generally sharper in the $S_{14}$ group compared to $S_{12}$, and it is particularly sharp for the center and 
higher sidebands in $S_{14}$.

There is also some disagreement between experiment and theory in the relationship
between the RABBITT phases for the various sidebands at small angles.  In particular, the experimental data for 
the lower sideband indicate a strong deviation from the other two, particularly in $S_{12,l}$ at the highest IR peak intensity of $12 \times 10^{11}\,$W/cm$^2$. Even though the model sensitivity of the theoretical predictions is generally small, it is visible particularly in the lower sideband of both groups.  The largest effect is seen in  $S_{12,l}$, where the presumably more accurate calculations (6st and 10st) predict $\Phi_R$ to start {\it below\/} the other two, while the experimental data have it starting {\it above} them.   
We currently do not have an explanation for this discrepancy. Since this problem is highly tractable with regard to the underlying theory, effectively electron scattering from a He$^+$ target, we are confident that our converged calculations are accurate within the assumed input parameters.
These calculations involve, of course, some idealizations whose accuracy regarding the physics is limited by the knowledge of and, subsequently, the ability to properly model all the intricacies in the actual experiment.

We finish, once again, with an illustration of the model sensitivity in the
results presented above.  Figure~\ref{fig:angleselected} exhibits the signal as a function of
the delay between the IR and the XUV train for both the angle-integrated cases as well as 
ejected-electron angles of $30^\circ$, $45^\circ$, and $60^\circ$.  In the 
$S_{12}$ and $S_{14}$ groups, one can clearly see the
phase of~$\pi$ predicted in the decomposition approximation between the center sideband and
the other two.  In $S_{16}$, on the other hand, that phase is there between the center and
the higher sideband, while the lower sideband is clearly different.

Finally, we note that it is not always the center sideband that shows the highest
contrast, in particular by reaching almost zero intensity for some delays.  Sometimes, the
higher sideband assumes this role.  In all cases, the lower sideband is the strongest, while
its oscillation amplitude relative to the average background is generally the smallest of 
the three. Nevertheless, the delay-dependence of all signals is sufficient to allow a highly
accurate fit to the expected signal and hence to extract the RABBITT phase.

\section{Conclusions and Outlook} \label{sec:Summary}
In this paper, we discussed the model sensitivity of our theoretical description of a recent
three-sideband RABBITT experiment in helium~\cite{Bharti2024}.  Except for the threshold sideband, which will be 
analyzed in a separate study to be carried out in the future, we found that $R$-matrix with
time dependence calculations in the simplest form of a 1-state model should be sufficient
to reproduce the principal characteristics of the process.  These include the dependence of 
the RABBITT phase on the laser intensities involved, as well as the angle dependence of the
phases in the three sidebands.  We found good agreement with the predictions of the 
decomposition approximation~\cite{Bharti2021} for two sideband groups, and significant 
aberrations in the two higher groups, where the weak XUV harmonics and the 
diminishing photo\-ionization cross section apparently lead to significant deviations in the lower sideband.
We note, however, that these deviations were indeed confirmed by experiment~\cite{Bharti2024}.

Looking back at the recent experiment and the present paper, we conclude that the 3-SB setup still contains a few surprises that warrant further investigation.  The helium target is promising in this respect, since several theoretical groups around the world should be able to carry out calculations of similar sophistication as the ones reported here.  The pulses used in our calculations are available on a numerical grid upon request. 

Of particular interest is the under-threshold RABBITT case.  With the present laser parameters, the first sideband just above threshold is coupled to the Rydberg spectrum, and hence the results become very sensitive to details of the model and the laser parameters.  We hope that the present paper will stimulate further investigations of this topic.


\begin{acknowledgments}
A.T.B., J.C.d.V, S.S., K.R.H., and K.B.\ acknowledge funding from the NSF through grant
\hbox{No.~PHY-2110023} as well as the  
Frontera Pathways allocation PHY-20028.
A.T.B.\ is grateful for funding through NSERC.
The calculations were performed on Stampede-2 and Frontera at the Texas Advanced Computing Center in Austin (TX).
The experimental part of this work was supported by the DFG-QUTIF program under
Project \hbox{No.~HA 8399/2-1} and \hbox{IMPRS-QD}.
\end{acknowledgments}

\bibliographystyle{apsrev4-1}

\end{document}